%% file: main.tex
\documentclass{lmcs}
\input{packages.tex}
\input{macros.tex}


\begin{document}

\title[Computing Distributed Knowledge]{On the Computation of Distributed Knowledge as the Greatest Lower Bound of Knowledge}
\thanks{This work has been partially supported by the ECOS-NORD project FACTS (C19M03) and the Minciencias project PROMUEVA (BPIN 2021000100160).}	

\author[S.~Quintero]{Santiago Quintero}[a]
\author[C.~Pinzon]{Carlos Pinz\'on}[b]
\author[S.~Ramirez]{Sergio Ram\'irez}[c]
\author[F.~Valencia]{Frank Valencia}[d]

\address{LIX, \'Ecole Polytechnique de Paris}	
\email{squinter@lix.polytechnique.fr}  

\address{Inria-LIX, \'Ecole Polytechnique de Paris}	
\email{carlos.pinzon@lix.polytechnique.fr}  

\address{Universidad EAFIT}	
\email{ssramirezr@eafit.edu.co}  

\address{CNRS-LIX, \'Ecole Polytechnique de Paris \and Pontificia Universidad Jareviana-Cali}	
\email{frank.valencia@lix.polytechnique.fr}  





\begin{abstract}
    Let $\Lat$ be a finite lattice and $\jhspace{\Lat}$ be the set of join endomorphisms of $\Lat$. 
    We consider the problem of given $\Lat$ and $f,g\in  \jhspace{\Lat}$, finding the greatest lower bound $f \meetp{\jhspace{\Lat}} g$ in the lattice $\jhspace{\Lat}$. (1) We show that if $\Lat$ is distributive, the problem can be solved in time $O(n)$ where $n=| \Lat |$. The previous upper bound was $O(n^2)$. (2) We provide new algorithms for arbitrary lattices and give experimental evidence that  they are significantly faster than the existing algorithm.  (3) We  characterize the standard notion of distributed knowledge of a group as the greatest lower bound of the join-endomorphisms representing the knowledge of each member of the group. (4) We show that deciding whether an agent has the distributed knowledge of two other agents can be computed in time $O(n^2)$ where $n$ is the size of the underlying set of states. (5) For the special case of $S5$ knowledge, we show that it  can be decided in time $O(n\invack{n})$ where $\invack{n}$ is  the inverse of the Ackermann function. 
\end{abstract}

\maketitle

\section{Introduction}

There is a long established tradition of using structures that involve a lattice $\Lat$ and its set of join-endomorphisms $\jhspace{\Lat}$ ordered pointwise. For instance, in \emph{modal algebras} \cite{bao-tarskip1-1951}, lattices appear as boolean algebras and their join-endomorphisms correspond, via duality, to \emph{box} modal operators. In \emph{concurrency theory}~\cite{milner-1989}, lattices have been used as orders of partial information and their join-endomorphisms represent either the spatial distribution or the perception of said information by the agents of a given system~\cite{knight:hal-00761116,guzman:hal-01257113}. In \emph{mathematical morphology}~\cite{bloch-mm-2007}, a well-established theory for the analysis and processing of geometrical structures founded upon lattice theory, join-endomorphisms correspond to one of its fundamental operations: \emph{dilations}.
In these and many other areas, lattices are used as rich abstract structures that capture the fundamental principles  of their domain of application.
 
Consequently, we believe that devising efficient algorithms  in the abstract  realm of lattice theory could  be useful as we may benefit from many representability results \cite{bao-tarskip1-1951,jonsson-canonicity-1994,gehrke-distlat-2004,gehrke-arblat-2001,dunn-posets-2005}  and 
identify general properties that can be exploited in the particular domain of application of the corresponding lattices. In fact, we shall use \emph{distributivity} and \emph{join-irreducibility} to reduce significantly the time and space needed to solve particular lattice problems. 

\subsection{Computing The Meet of Join-Endomorphism}
In this paper we shall focus on algorithms for the following \emph{maximization} problem: \emph{Given a lattice $\Lat$ of size $n$ and two join-endomorphism $f,g \in \jhspace{\Lat}$, find the  greatest lower bound  $h= f \meetp{\jhspace{\Lat}} g$} (i.e. the greatest join-endomorphism $h \in \jhspace{\Lat}$ below $f$ and $g$). Notice that the input lattice is $\Lat$ not $\jhspace{\Lat}$.
Simply taking $h(a)=f(a) \meetp{\Lat} g(a)$ for all $a \in \Lat$ does not work because the resulting $h$ may not even be a join-endomorphism. Previous lower bounds for solving this problem are $O(n^3)$ for arbitrary lattices and $O(n^2)$ for distributive lattices~\cite{quintero-ramics-2020}. By using abstract properties of lattices we will show that this problem can actually be solved in $O(n)$ for distributive lattices. 

Furthermore, we will exploit properties of join-endomorphisms to develop heuristics for this maximization problem. We will provide significant experimental evidence from randomly generated lattices that these heuristics improve considerably the time performance of previous algorithms.  In fact, the reader will be able to validate the performance results by using a platform available at \url{https://caph1993.github.io/GMeetMono/}.

\subsection{Computing Distributed Knowledge}

\emph{Distributed knowledge}~\cite{HalpernM90} corresponds to knowledge that is distributed among the members of a group, without any of its members necessarily having it.  This notion can be used to analyse the implications of the knowledge of a community if its members were to combine their knowledge, hence its importance.

We will show that the distributed knowledge of a group can be seen as the meet of the join-endomorphisms representing the knowledge of its members. To do this we use the standard  structures in economics for multi-agent knowledge~\cite{samet-agreeing-2010} which involve a set of \emph{states (or worlds)} $\Stateset$ and a \emph{knowledge operator (function)} $\Kfun{i}:\pset{\Stateset}\to\pset
{\Stateset}$ describing the events, represented as subsets of $\Stateset$, that an agent $i$ knows. The event of  $i$ knowing the event $E$ is  $\Kfunapp{i}{E} = \{ \state \in \Stateset \mid \Relapp{i}{\state} \subseteq E \}$ where $\Rel{i} \subseteq \Stateset^2$  is the accessibility relation of $i$ and $\Relapp{i}{\state} = \{ \state' \mid (\state,\state')\in\Rel{i}\}.$  The event of having distributed knowledge of $E$ by $i$ and $j$ is $\DKfunapp{i,j}{E} = \{ \state \in \Stateset \mid \Relapp{i}{\state} \cap \Relapp{j}{\state} \subseteq E \}$~\cite{fagin1995reasoning}.  

It turns out that knowledge operators are join-endomorphisms of the lattice $\Lat = (\pset{\Stateset},\supseteq)$.  Intuitively, the lower an agent $i$ (its knowledge function) is placed in $\jhspace{\Lat}$, the ``wiser" (or more knowledgeable) the agent is. {We will show that $\DKfun{i,j}= \Kfun{i}\meetp{\jhspace{\Lat}}\Kfun{j}.$}  This means $\DKfun{i,j}$ can  be viewed as the \emph{least knowledgeable agent wiser than both $i$ and $j$}.

We also consider the following decision problem: \emph{Given the knowledge of agents $i$, $j$, and $m$, decide whether $m$ has the distributed knowledge of $i$ and $j$, i.e. whether $\Kfun{m}=\DKfun{i,j}$}. The knowledge of an agent $k$ can be represented by $\Kfun{k}:\pset{\Stateset}\to\pset
{\Stateset}$. If available it can also be represented, exponentially more succinctly,  by $\Rel{k} \subseteq \Stateset^2$. In the first case the problem reduces to checking whether $\Kfun{m}= \Kfun{i}\meetp{\jhspace{\Lat}}\Kfun{j}$. In the second the problem reduces to  $\Rel{m}=\Rel{i}\cap\Rel{j}$ and this can be done in $O(n^2)$ where $n=|\Stateset|$.

Nevertheless, we show that even without the accessibility relations,  if the only inputs are the knowledge operators, represented as arrays, the problem can be still be solved in $O(n^2)$. We obtain this result using tools from lattice theory to exponentially reduce the number of tests on the knowledge operators (arrays) needed to decide the problem.

Furthermore, if the inputs are the accessibility relations and they are equivalences (hence they can be represented as partitions), we show that the problem can be solved basically in \emph{linear time}: More precisely, in $O(n\invack{n})$ where  $\invack{n}$ 
is an extremely slow growing function; the inverse of the Ackermann function.  It is worth noticing that if accessibility relations can be represented as partitions, the structures are known as Aumann structures~\cite{aumann-disagree-1976} and they characterize a standard notion of knowledge called  $S5$~\cite{fagin1995reasoning}. 

\subsection*{Efficient Intersection of Partitions}
To prove the above-mentioned $O(n\invack{n})$ bound we show a new result of independent interest using a Disjoint-Set data structure~\cite{galler1964improved}. Recall that the intersection of two partitions $X$ and $Y$ of a set $S$ is the partition $Z$ of $S$ such that every $a,b \in S$, $a \sim_Z b$ iff both $a \sim_X b$ and $a \sim_Y b$, where $a \sim_W b$ means that $a$ and $b$ are in the same block of a partition $W$ of $S$. We shall prove that using Disjoint-Set data structures, the intersection of two \emph{partitions} of a set of size $n$ can be computed in $O(n\invack{n})$. To our knowledge this is a new result for Disjoint-Set data structures. 

The above intersection result may have applications beyond \emph{knowledge}, particularly in graph theory where the Disjoint-Set data structure is widely used. For example, our result implies the following graph theoretical problem can be computed in $O(n\invack{n})$ using a Disjoint-Set data structure:
\emph{Given two undirected graphs $G_1$ and $G_2$ with the same $n$ nodes, find an undirected graph $G_3$ such that two nodes are connected in it iff they are connected in both $G_1$ and $G_2$.}

\subsection{Main Results and Organization.} 
In summary, the main contributions of this paper are the following:
\begin{enumerate}

\item We prove that for \emph{distributive} lattices of size $n$, the meet of join-endomorphisms can be computed in time $O(n)$. Previous upper bound was $O(n^2)$.

\item We present new algorithms for computing the meet of join-endomorphisms for \emph{arbitrary} lattices and provide experimental evidence on randomly generated lattices and join-endomorphisms  that they significantly outperform the previous algorithm.   We shall also provide the implementation code in Pynthon of these algorithms as well as the code for the generation of random lattices. We also provide an interactive interface where the reader can run experiments and validate the performance results. This platform is available at \url{https://caph1993.github.io/GMeetMono/}.\label{cont:algo}
 
\item We show that distributed knowledge of a given group can be viewed as the meet of the join-endomorphisms representing the knowledge of each member of the group. 

\item We show that the problem of whether an agent has the distributed knowledge of two other can be decided in time $O(n^2)$ where $n=|\Stateset|$.\label{cont:dkp}

\item If the agents' knowledge can be represented as partitions, the problem in (\ref{cont:dkp}) can be decided in $O(n\invack{n})$. To obtain this we provide a procedure, interesting in its own right, that computes the intersection of two \emph{partitions} of a set of size $n$ in $O(n\invack{n})$.
\end{enumerate}

\paragraph{Other Results} The above results are given in Sections~\ref{sec:meet-joinendo} and~\ref{sec:AK}. For conducting our study, in the intermediate sections (Sections~\ref{sec:j-endo-irred} and~\ref{sec:app-know}) we will adapt some representation and duality results (e.g. J\'onsson-Tarski duality~\cite{bao-tarskip2-1952}) to our structures. Some of these results are part of the folklore in lattice theory but for completeness we provide simple proofs of them.

\paragraph{Note.} This paper has been submitted to the LMCS special issue of RAMiCS 2021 as the extended version of \cite{quintero-ramics-2021}  with full proofs of each result as well as with the new theoretical and experimental material mentioned in (\ref{cont:algo}) in the above-listed mains results.

\section{Notation, Definitions and Elementary Facts}
\label{sec:back}

We list facts and notation used throughout the paper. We index joins, meets, and order relations with their corresponding poset but often omit the index when it is clear from the context. 

\paragraph{{\bf Partially Ordered Sets and Lattices.}}
A poset $\C$ is a \emph{lattice} iff each finite nonempty subset of $\Con$ has a supremum and infimum in $\Con$. It is a \emph{complete lattice} iff each subset of $\Con$ has a supremum and infimum in $\Con$. A poset $\C$ is \emph{distributive} iff for every $a,b,c \in \Con$, $a \join ( b \meet c) = (a \join b) \meet (a \join c)$. We write $a \| b$ to denote that $a$ and $b$ are incomparable in the underlying poset.
A \emph{lattice of sets} is a set of sets ordered by inclusion and closed under finite unions and intersections. A \emph{powerset lattice} is a lattice of sets that includes all the subsets of its top element.

\begin{defi}[Downsets, Covers, Join-irreducibility~\cite{davey2002introduction}]
\label{def:downset}
Let $\Lat$ be a lattice and $a,b \in \Lat$. We say $b$ is covered by $a$, written $b \cov a$, if $b \cl a$ and there is no $c \in \Lat$ s.t., $b \cl c \cl a$.
The  \emph{down-set} \emph{(up-set)} of $a$ is $\downset{a} \defsymbol \{b \in \Lat \mid b \cleq a\}$ $(\upset{a} \defsymbol \{b \in \Lat \mid b \cgeq a\})$, and the set of elements \emph{covered} by $a$ is $\covers{a} \defsymbol \{ b \mid b \cov a\}$.
An element $c \in \Lat$ is said to be \emph{join-irreducible} if 
$c = a \join b$ implies $c = a$ or $c = b$.
If $\Lat$ is finite, $c$ is \emph{join-irreducible} if $|\covers{c}|=1$. The set of all join-irreducible elements of $\Lat$ is $\ijoin{\Lat}$ and $\jdown{c} \defsymbol \downset{c} \cap \ijoin{\Lat}$.
\end{defi}

\paragraph{{\bf Posets of maps.}} A map $f: X \to Y$ where $X$ and $Y$ are posets is
\emph{monotonic (or order-preserving)} if $a \cleqp{X} b$ implies $f(a) \cleqp{Y} f(b)$ for every $a,b \in X$. We say that $f$ \emph{preserves} the join of $S \subseteq X$  iff
$f(\bigjoin S) = \bigjoin \{ f(c)\mid c \in S \}$.
A \emph{self-map} on $X$ is a function $f: X \to X$. 
If $X$ and $Y$ are posets, we define $\Fvar$ as the poset of all functions from $X$ to $Y$. We use $\langle X \to Y \rangle$ to denote the poset of monotonic functions of $\Fvar$.
The functions in $\Fvar$ are ordered pointwise: i.e. $f \cleqp{\Fvar} g$ iff $f(a) \cleqp{Y} g(a)$ for every $a\in X$.

\begin{defi}[Join-endomorphisms and $\jhspace{\Lat}$]
\label{def:join-endo}
Let $\Lat$ be a lattice.  We say that a self-map is a 
\emph{(bottom preserving) join-endomorphism} iff it preserves the join of every finite
subset of $\Lat$. Define $\jhspace{\Lat}$ as the set of all
join-endomorphisms  of $\Lat$. Furthermore,  given $f,g \in \jhspace{\Lat}$,
define $f \hleq g$ iff $f(a) \cleq g(a)$ for every $a\in\Lat$.
\end{defi}

The following are properties that we shall use throughout the paper.

\begin{prop}[\cite{gratzer-latjoinend-1958,davey2002introduction}]
\label{prop:jendo-proprts}
Let $\Lat$ be a lattice.
\begin{enumerate}[label=\textbf{P.\arabic*},ref=P.\arabic*]
    \item $f \in \jhspace{\Lat}$ iff $f(\bot) =\bot$ and $f(a \join b) = f(a) \join f(b)$ for all $a,b\in\Lat$.
    \label{prop:jendo-proprts1}
    \item If $f \in \jhspace{\Lat}$ then $f$ is monotonic.
    \label{prop:jendo-proprts2}
    \item If $\Lat$ is a complete lattice, then $\jhspace{\Lat}$ is a complete lattice.
    \label{prop:jendo-proprts7}
    \item $\jhspace{\Lat}$ is a complete distributive lattice iff $\Lat$ is a complete distributive lattice.
    \label{prop:jendo-proprts3}
    \item If $\Lat$ is finite and distributive, $\jhspace{\Lat} \cong \imonspace{\Lat}$.
    \label{prop:jendo-proprts4}
    \item If $\Lat$ is a finite lattice,  $e = \bigjoinp{\Lat} \{c \in \ijoin{\Lat} \mid c \cleq e\}$ for every $e\in\Lat$.
    \label{prop:jendo-proprts5}
    \item If $\Lat$ is finite and distributive, $f \in \jhspace{\Lat}$ iff $(\forall e \in \Lat)\ f(e) = \bigjoin \{ f(e')\ |\ e' \in \jdown{e}\}$.
    \label{prop:jendo-proprts6}
\end{enumerate}
\end{prop}

We shall use these posets in our examples:  $\bar{\mathbf{n}}$ is $\{1,\ldots,n\}$ with the order $x \cleq y$ iff $x=y$ and $\M_n \defsymbol  (\bar{\mathbf{n}}_\bot)^\top$ is the lattice that results from adding  a top and bottom to $\bar{\mathbf{n}}$.

\section{Computing the Meet of Join-Endomorphisms}
\label{sec:meet-joinendo}

Join-endomorphisms and their meet arise as fundamental computational operations in computer science.
We therefore believe that the problem of computing these operations in the abstract realm of lattice theory is a relevant issue: We may identify general properties that can be exploited in all instances of these lattices.  

In this section, we address the problem of computing the meet of join-endomorphisms.
Let us consider the following \emph{maximization} problem.

\begin{prob}
\label{prob:meet}
Given a lattice $\Lat$ of size $n$ and two join-endomorphisms $f,g:\Lat \to \Lat$, find the \emph{greatest} join-endomorphism $h:\Lat \to \Lat$ below both $f$ and $g$: i.e. $h=f \meetp{\jhspace{\Lat}} g$.
\end{prob}

Notice that the lattice $\jhspace{\Lat}$, which could be exponentially bigger than $\Lat$~\cite{quintero-ramics-2020}, is not an input to the problem above.
It may not be immediate how to find $h$; e.g. see the endomorphism $h$ in Figure~\ref{fig:difficult-small} for a small lattice of four elements. 
A \emph{naive approach} to find $f \meetp{\jhspace{\Lat}} g$ could be  to attempt to compute it pointwise  by taking $h(a) = f(a) \meet_\Lat g(a)$ for every $a \in \Lat$.
Nevertheless, the somewhat appealing equation
\begin{equation}\label{eq:naive-eq:1}
\left(f \meetp{\jhspace{\Lat}} g\right)(a) = f(a) \meetp{\Lat} g(a)
\end{equation}
\emph{does not} hold in general, as illustrated in the lattices $\M_2$ and $\M_3$ in Figure~\ref{fig:meet-pw} and Figure~\ref{ex:m3}. 

A general approach in~\cite{quintero-ramics-2020} for arbitrary lattices shows how to find $h$ in Problem~\ref{prob:meet} by successive approximations $\dapprox^0  \cg\dapprox^1 \cg \cdots  \cg \dapprox^i$, starting with some self-map $\dapprox^0$ known to be smaller than both $f$ and $g$, and greater than $h$; while keeping the invariant $\dapprox^{i} \cgeq h$.
The starting point is the naive approach above: $\dapprox^{0}(a) = f(a)\meet g(a)$ for all $a\in \Lat.$
The approach computes decreasing upper bounds of $h$ by correcting in $\dapprox^{i}$  the image under $\dapprox^{i-1}$ of some values $b, c, b \join c$  violating the property \( \dapprox^{i-1}(b)
\join\dapprox^{i-1}(c)=\dapprox^{i-1}(b\join c).\) The correction satisfies $\dapprox^{i-1} \cg \dapprox^i $ and
maintains the invariant  $\dapprox^{i} \cgeq h$. This approach eventually finds $h$ in $O(n^3)$ basic lattice operations (binary meets and joins). 

For the sake of the presentation, we approach the above problem for distributive and arbitrary lattices separately.

\subsection{Algorithms for Distributive Lattices}
\label{sec:dist-lat}

Recall that in finite distributive lattices, and more generally in co-Heyting algebras \cite{tarski-closed-elem-1946}, the subtraction operator $\sop$ is uniquely determined by the \emph{Galois connection} $b \cgeq c \sop a$ iff $a \join b \cgeq c$. Based on the following proposition, it was shown in~\cite{quintero-ramics-2020} that if the only basic operations are joins or meets, $h$ can be computed in $O(n^3)$ of them. If we also allow subtraction as a basic operation, the bound can be improved to $O(n^2)$. 
\begin{prop}[\cite{quintero-ramics-2020}]
\label{prop:comp-algo}
Let $\Lat$ be a finite distributive lattice. Let $h = f \meetp{\jhspace{\Lat}} g$. Then
\begin{enumerate}
  \item \(h(c) = \bigmeetp{\Lat}\left\{ f(a) \join g(b) \mid a \join b \cgeq c \right\}\), and
  \item \(h(c) = \bigmeetp{\Lat}\left\{ f(a) \join g(c \sop a) \mid a \in \downset{c} \right\}.\)
\end{enumerate} 
where $c \sop a \defsymbol \bigmeetp{\Lat}\{ e \mid a \join e \cgeq c \}.$
\end{prop}

\begin{figure}[t]
    \centering
    \begin{subfigure}[b]{0.3\textwidth}
    \centering
    \begin{tikzpicture}[scale=0.3,>=stealth]
      \tikzstyle{every node}=[font=\scriptsize]
      \node[shape = circle, draw] (A) at (0,-5)   {$\bot$};
      \node[shape = circle, draw] (B) at (-5,0)   {$1$};
      \node[shape = circle, draw] (C) at (5,0)    {$2$};
      \node[shape = circle, draw] (D) at (0,5)    {$\top$};
    
      \draw (A) to node {} (B);
      \draw (A) to node {} (C);
      \draw (B) to node {} (D);
      \draw (C) to node {} (D);
      \draw [above,blue,->,bend left,dotted,thick]     (B) to node {} (C);
      \draw [below,blue,->,bend left,dotted,thick]     (C) to node {} (B);
      \draw [below,blue,->,loop left,dotted,thick]     (A) to node {} (A);
      \draw [below,blue,->,loop left,dotted,thick]     (D) to node {} (D);
      \draw [below,red,->,loop right,thick]            (D) to node {} (D);
      \draw [left,red,->,bend left,thick]              (B) to node {} (D);
      \draw [below,red,->,loop right,thick]            (A) to node {} (A);
      \draw [below,red,->,loop below,thick]            (C) to node {} (C);
    
      \draw [below,teal,dashed,loop below,->,thick]  (A) to node {} (A);
      \draw [right,teal,dashed,bend left,->,thick]   (D) to node {} (C);
      \draw [above,teal,dashed,bend left,->,thick]   (C) to node {} (A);
      \draw [above,teal,dashed,->,thick]             (B) to node {} (C);
    \end{tikzpicture}
    \caption{{\color{blue} $f : \dottedarrow$}, {\color{red} $g {:} \rightarrow$}, {\color{teal} $h {:} \dashrightarrow$}}
    \label{fig:difficult-small}
    \end{subfigure}
    ~
    \begin{subfigure}[b]{0.3\textwidth}
    \centering
    \begin{tikzpicture}[scale=0.3,>=stealth]
      \tikzstyle{every node}=[font=\scriptsize]
      \node[shape = circle, draw] (A) at (0,-5)   {$\bot$};
      \node[shape = circle, draw] (B) at (-5,0)   {$1$};
      \node[shape = circle, draw] (C) at (5,0)    {$2$};
      \node[shape = circle, draw] (D) at (0,5)    {$\top$};
    
      \draw (A) to node {} (B);
      \draw (A) to node {} (C);
      \draw (B) to node {} (D);
      \draw (C) to node {} (D);
      \draw [above,blue,->,bend left,dotted,thick]     (B) to node {} (C);
      \draw [below,blue,->,bend left,dotted,thick]     (C) to node {} (B);
      \draw [below,blue,->,loop left,dotted,thick]     (A) to node {} (A);
      \draw [below,blue,->,loop left,dotted,thick]     (D) to node {} (D);
      \draw [below,red,->,loop right,thick]            (D) to node {} (D);
      \draw [left,red,->,bend left,thick]              (B) to node {} (D);
      \draw [below,red,->,loop right,thick]            (A) to node {} (A);
      \draw [below,red,->,loop above,thick]            (C) to node {} (C);
    
      \draw [below,teal,dashed,loop below,->,thick]  (A) to node {} (A);
      \draw [above,teal,dashed,loop above,->,thick]  (D) to node {} (D);
      \draw [right,teal,dashed,bend left,->,thick]   (C) to node {} (A);
      \draw [above,teal,dashed,->,thick]             (B) to node {} (C);
    \end{tikzpicture}
    \caption{{\color{blue} $f : \dottedarrow$}, {\color{red} $g {:} \rightarrow$}, {\color{teal} $h {:} \dashrightarrow$}}
    \label{fig:meet-pw}
    \end{subfigure}
    ~
    \begin{subfigure}[b]{0.3\textwidth}
    \centering
    \begin{tikzpicture}[scale=0.3,>=stealth]
      \tikzstyle{every node}=[font=\scriptsize]
      \node [shape = circle, draw] (A) at (0,-5)   {$\bot$};
      \node [shape = circle, draw] (B) at (-5,0)   {$1$};
      \node [shape = circle, draw] (C) at (0,0)    {$2$};
      \node [shape = circle, draw] (D) at (5,0)    {$3$};
      \node [shape = circle, draw] (E) at (0,5)    {$\top$};
    
      \draw (A) to node {} (B);
      \draw (A) to node {} (C);
      \draw (A) to node {} (D);
      \draw (B) to node {} (E);
      \draw (C) to node {} (E);
      \draw (D) to node {} (E);
    
      \draw [left,loop above,blue,->,dotted,thick]     (B) to node {} (B);
      \draw [above,blue,->,bend left,dotted,thick]     (C) to node {} (D);
      \draw [below,blue,->,bend left,dotted,thick]     (D) to node {} (C);
      \draw [below,loop below,blue,->,dotted,thick]    (A) to node {} (A);
      \draw [above,loop above,blue,->,dotted,thick]    (E) to node {} (E);
      \draw [above,loop right,red,->,thick]            (E) to node {} (E);
      \draw [left,loop right,red,->,thick]             (C) to node {} (C);
      \draw [above,red,->,bend left,thick]             (B) to node {} (E);
      \draw [right,loop above,red,->,thick]            (D) to node {} (D);
      \draw [below,loop left,red,->,thick]             (A) to node {} (A);
    
      \draw [above,loop right,teal,->,dashed,thick]  (A) to node {} (A);
      \draw [left,teal,->,bend right,dashed,thick]   (C) to node {} (A);
      \draw [right,teal,->,bend left,dashed,thick]   (D) to node {} (A);
      \draw [below,loop below,teal,->,dashed,thick]  (B) to node {} (B);
    \end{tikzpicture}
    \caption{{\color{blue} $f : \dottedarrow$}, {\color{red} $g {:} \rightarrow$}, {\color{teal} $h {:} \dashrightarrow$}}
    \label{ex:m3}
  \end{subfigure}
\caption{(a) $h = f \meetp{\jhspace{\Lat}} g$.
(b) $h(a)\defsymbol f(a) \meet g(a)$ for $a \in \M_2$ is not in $\jhspace{\M_2}$: $h(1 \join 2)\neq h(1) \join h(2)$.
(c) Any $h:\M_3 \to \M_3$ s.t. $h(a) = f(a) \meet g(a)$ for $a \in \ijoin{\M_3}$ is not in $\jhspace{\M_3}$: $h(\top) = h(1 \join 2) = h(1) \join h(2) = 1 \neq \bot = h(2) \join h(3) = h(2 \join 3) = h(\top).$}
\label{fig:mn-lat}
\end{figure}

Nevertheless, it turns out that we can partly use Equation~\ref{eq:naive-eq:1} to obtain a better upper bound.
The following lemma states that Equation~\ref{eq:naive-eq:1} holds if $\Lat$ is distributive and $a \in \ijoin{\Lat}.$

\begin{lem}
\label{lem:jimeet}
Let $\Lat$ be a finite distributive lattice and $f,g \in \jhspace{\Lat}$. Then
\[\left( f \meetp{\jhspace{\Lat}} g \right)(a) = f(a) \meetp{\Lat} g(a)\]
holds for every $a \in \ijoin{\Lat}$.
\end{lem}

\begin{proof}
From Proposition~\ref{prop:comp-algo},
\(({f \meetp{\jhspace{\Lat}} g})(a) =
\bigmeet \left\{ f(a') \join g(a \sop a')\ |\ a' \in \downset{a} \right\}.\)
Note that since  $a \in \ijoin{\Lat}$ if $a' \in \downset{a}$ then $a \sop a' = a$ when $a \neq a'$, and $a \sop a' = \bot$
when $a = a'$. Then,
\[
\begin{array}{lcl}
     \{ f(a') \join g(a \sop a')\ |\ a' \in \downset{a}\} & = & \{ f(a') \join g(a \sop a')\ |\ a' \cl a\} \cup \{f(a) \join g(\bot)\} \\
     & = & \{ f(a') \join g(a)\ |\ a' \cl a\} \cup \{f(a)\} \\
     & = & \{ f(a') \join g(a)\ |\ \bot \cl a' \cl a\} \cup \{f(a), g(a)\}.
\end{array}
\]
By absorption, we know that $(f(a') \join g(a)) \meet g(a) = g(a)$. Finally, using properties of $\meet$, $(f \meetp{\jhspace{\Lat}} g)(a) = \bigmeet (\{ f(a') \join g(a)\ |\ \bot \cl a' \cl a\} \cup \{f(a), g(a)\}) = \bigmeet \{ f(a') \join g(a)\ |\ \bot \cl a' \cl a\} \meet f(a) \meet g(a) = f(a) \meet g(a)$.
\end{proof}

It is worth noting the Lemma~\ref{lem:jimeet} may not hold for non-distributive lattices. This is illustrated in Figure~\ref{ex:m3} with the archetypal non-distributive lattice $\M_3$. Suppose that $f$ and $g$ are given as in Figure~\ref{ex:m3}.
Let $h = f \meetp{\jhspace{\Lat}} g$ with  $h(a) = f(a) \meet g(a)$ for all $a \in \{1,2,3\} =\ijoin{\M_3}$. Since $h$ is a join-endomorphism, we would have $h(\top) = h(1 \join 2) = h(1) \join h(2) = 1 \neq \bot = h(2) \join h(3) = h(2 \join 3) = h(\top)$, a contradiction.

Lemma~\ref{lem:jimeet} and Property~\ref{prop:jendo-proprts6} lead us to the following characterization of meets over $\jhspace{\Lat}$.

\begin{thm}\label{thm:meet-ch} Let $\Lat$ be a finite distributive lattice and $f,g \in \jhspace{\Lat}$.   Then $h = f \meetp{\jhspace{\Lat}} g$ iff $h$ satisfies 
\begin{equation}\label{eq:h}
h(a) =
\begin{cases}
f(a) \meet_\Lat g(a) 		\quad & 	  \mbox{ if } a \in \ijoin{\Lat}  \mbox{ or } a = \bot \\
h(b)  \join_\Lat  h(c)           	\quad &       \mbox{ if } b,c \in \covers{a} \mbox{ with } b\neq c\\
\end{cases}
\end{equation}
\end{thm}

\begin{proof}
    The only-if direction follows from Lemma~\ref{lem:jimeet} and~\ref{prop:jendo-proprts6}.  For the if-direction, suppose that $h$ satisfies Equation~\ref{eq:h}. If $h \in \jhspace{\Lat}$ the result
    follows from Lemma~\ref{lem:jimeet} and \ref{prop:jendo-proprts6}. To prove $h \in \jhspace{\Lat}$ from \ref{prop:jendo-proprts6} it suffices to show  \inlineeq{h(e) = \bigjoin \{ h(e') \mid e' \in \jdown{e}\}}\label{eq:prop-ind-h} for every $e \in \Lat$.  From Equation~\ref{eq:h} and since $f$ and $g$ are monotonic,  $h$ is monotonic.  If $e \in \ijoin{\Lat}$ then $h(e') \cleq h(e)$ for every $e' \in \jdown{e}$. Therefore, $\bigjoin \{ h(e') \mid e' \in \jdown{e}\} = h(e)$.
    If $e \not\in \ijoin{\Lat}$, we proceed by induction.  Assume Equation~\ref{eq:prop-ind-h} holds for all $a \in \covers{e}$. By definition, $h(e) = h(b) \join h(c)$ for any $b,c \in \covers{e}$ with $b \neq c$. Then, we have $h(b) = \bigjoin \{ h(e') \mid e' \in \jdown{b}\}$ and $h(c) = \bigjoin \{ h(e') \mid e' \in \jdown{c}\}$. Notice that $e' \in \jdown{b}$ or $e' \in \jdown{c}$ iff $e' \in \jdown{(b \join c)}$, since $\Lat$ is distributive.
    Thus, $h(e) = h(b) \join h(c) = \bigjoin \{ h(e') \mid  e' \in \jdown{(b \join c)}\} = \bigjoin \{ h(e') \mid  e' \in \jdown{e}\}$ as wanted.
\end{proof}

We conclude this section by stating the time complexity $O(n)$ to compute $h$ in the above theorem. As in \cite{quintero-ramics-2020},  the time complexity  is 
determined by the number of basic binary lattice operations (i.e. meets and joins) performed during execution. 

\begin{cor}  
\label{thm:jecomplexity}
Given a distributive lattice $\Lat$ of size $n$, and functions
$f, g \in \jhspace{L}$, the function $h = f \meetp{\jhspace{\Lat}} g$ can be 
computed in $O(n)$ binary lattice operations.
\end{cor}

\begin{proof}
If $a \in \ijoin{L}$ then from 
Theorem~\ref{thm:meet-ch}, $h(a)$ can be computed as $f(a) \meet g(a)$. If $a =\bot$ then $h(a)$ is $\bot$. 
If $a \notin \ijoin{\Lat}$ and $a \neq \bot$, we pick any $b, c \in \covers{a}$ such that $b \neq c$
and compute $h(a)$ recursively as $h(b) \join h(c)$ by Theorem~\ref{thm:meet-ch}.
We can use a lookup table to keep track of the values of $a \in \Lat$ for which $h(a)$ has been computed, starting with all $a \in \ijoin{\Lat}$.
Since $h(a)$ is only computed once for each $a \in \Lat$, either as a meet for elements in $\ijoin{\Lat}$ or as a join otherwise,
we only perform $n$ binary lattice operations.
\end{proof}

\subsubsection{Experimental Results}
Now we present some experimental results comparing the average runtime between the previous algorithm in \cite{quintero-ramics-2020} based on Proposition~\ref{prop:comp-algo}, referred to as \DMeet{}, and the proposed algorithm in Theorem~\ref{thm:meet-ch}, called \DMeetPlus{}.
\begin{figure}[t]
    \centering
    \begin{subfigure}[b]{0.48\textwidth}
      \centering
      \includegraphics[width=\textwidth]{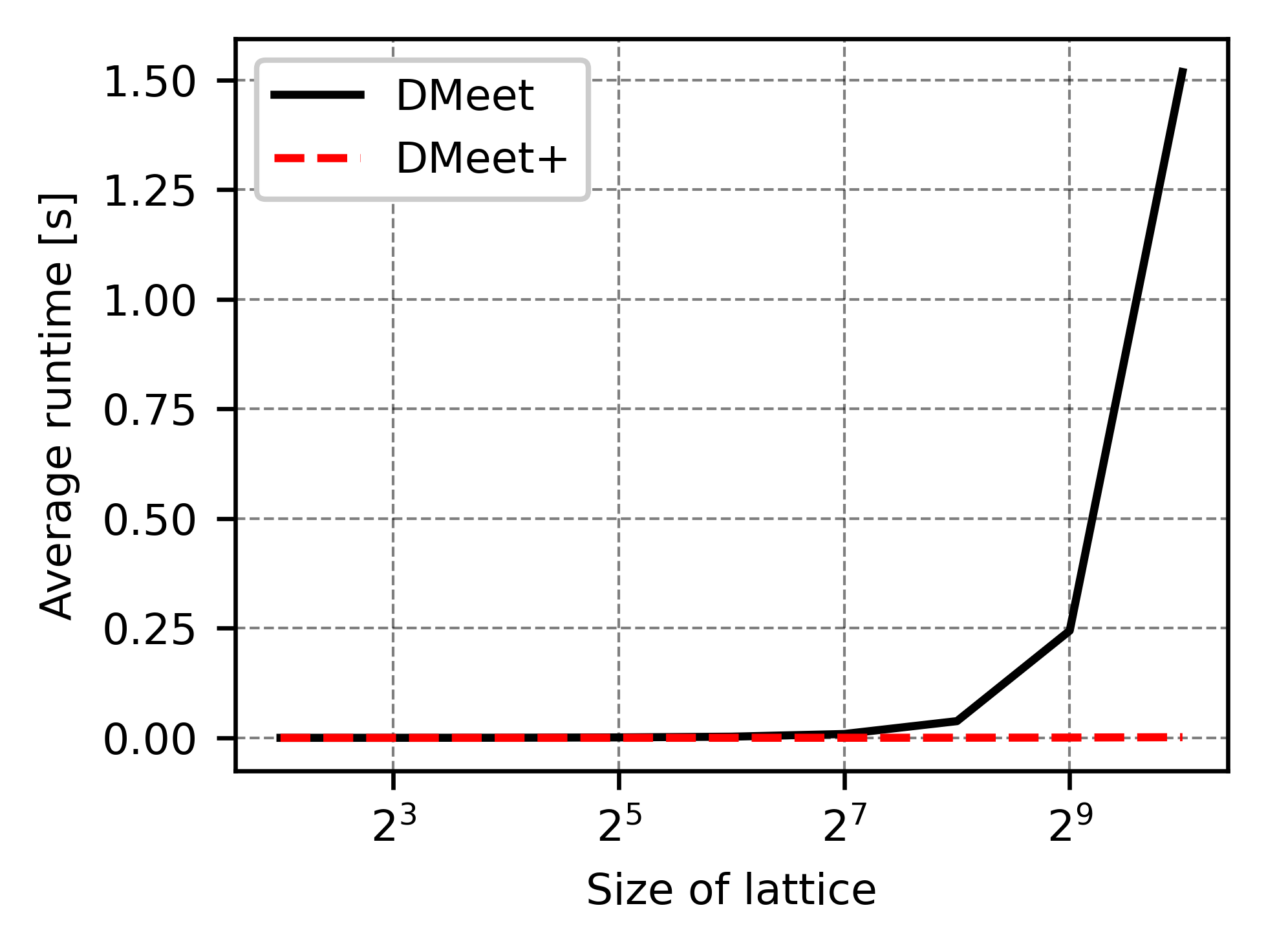}
      \caption{Powerset lattices.}
      \label{fig:dmeet-powerset}
    \end{subfigure}
    ~
    \begin{subfigure}[b]{0.49\textwidth}
      \centering
      \includegraphics[width=\textwidth]{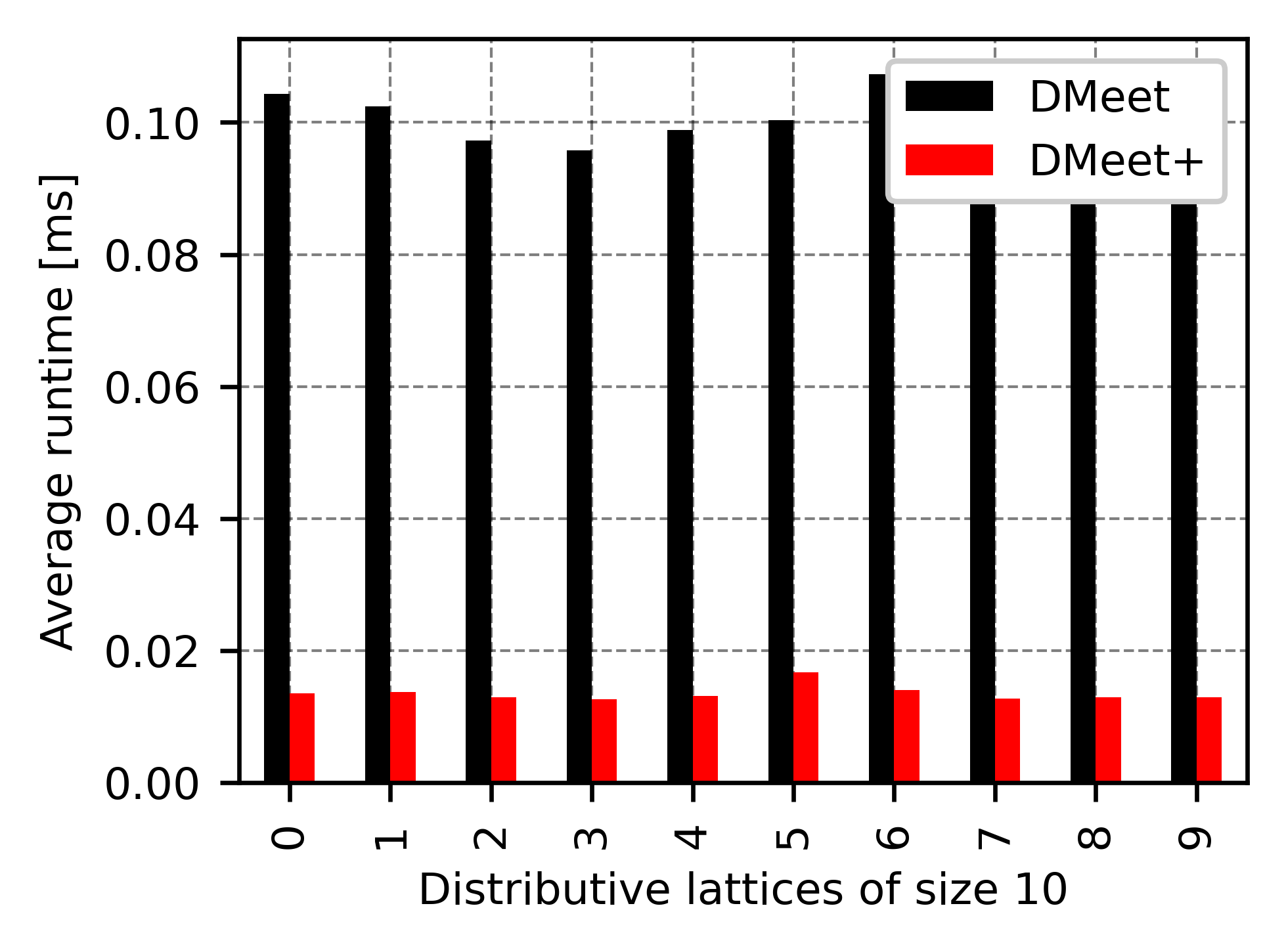}
      \caption{Arbitrary distributive lattices.}
      \label{fig:dmeet-dist}
    \end{subfigure}
\caption{Comparison between an implementation of Proposition~\ref{prop:comp-algo} (\DMeet{})
and Theorem~\ref{thm:meet-ch} (\DMeetPlus{}).}
\label{fig:dmeet-experiments}
\end{figure}

Figure~\ref{fig:dmeet-experiments} shows the average runtime of each algorithm, from 100 runs with a random pair of join-endomorphisms. For Figure~\ref{fig:dmeet-powerset}, we compared each algorithm against powerset lattices of sizes between $2^2$ and $2^{10}.$
For Figure~\ref{fig:dmeet-dist}, 10 random distributive lattices of size 10 were selected. In both cases, all binary lattice
operation are guaranteed a complexity in $O(1)$ to showcase the quadratic nature of \DMeet{} compared to the linear growth of \DMeetPlus{}.
The time reduction from \DMeet{} to \DMeetPlus{} is also reflected in a reduction on the number of $\join$ and $\meet$ operations performed as illustrated in Table~\ref{tab:dmeet-experiments}.
For  \DMeetPlus{}, given a distributive lattice $\Lat$ of size $n$, \#$\meet = |\ijoin{\Lat}|$ and \#$\join = |\Lat| - |\ijoin{\Lat}| - 1$ ($\bot$ is directly mapped to $\bot$). 

\begin{table}
\begin{center}
\setlength{\tabcolsep}{10pt}
\begin{tabular}{ l r r r r r r }
    \toprule
    {} & \DMeet{} & \DMeetPlus{} & \DMeet & \DMeetPlus{} & \DMeet{} & \DMeetPlus{} \\
    Size &  Time [s] &  Time [s] &  \#$\join$ &  \#$\join$  &  \#$\meet$  &  \#$\meet$ \\
    \midrule
    16    &    0.000246 &   0.000024 &          81 &         11 &          81 &          4 \\
    32    &    0.000971 &   0.000059 &         243 &         26 &         243 &          5 \\
    64    &    0.002659 &   0.000094 &         729 &         57 &         729 &          6 \\
    128   &    0.008735 &   0.000163 &        2187 &        120 &        2187 &          7 \\
    256   &    0.038086 &   0.000302 &        6561 &        247 &        6561 &          8 \\
    512   &    0.244304 &   0.000645 &       19683 &        502 &       19683 &          9 \\
    1024  &    1.518173 &   0.001468 &       59049 &       1013 &       59049 &         10 \\
    \bottomrule
\end{tabular}
\caption{Average runtime in seconds over powerset lattices. Number of $\join$ and $\meet$ operations performed for each algorithm.}
\label{tab:dmeet-experiments}
\end{center}
\end{table}

\subsection{Algorithms for Arbitrary Lattices}
\label{sec:new-section-gmeet}

The \DMeetPlus{} algorithm, introduced in Section~\ref{sec:dist-lat}, computes the meet of join endomorphisms on distributive lattices in $O(n)$.
This section explores algorithms for computing the meet of join endomorphisms on arbitrary lattices, not necessarily distributive.
The best known algorithm for this task is \GMeetPlus{}introduced in~\cite{quintero-ramics-2020}, which is based on successive approximations (as described at the beginning of Section~\ref{sec:meet-joinendo}) and has a complexity of $O(n^3)$.
This section presents alternative algorithms for the same task, each with its proof of correctness and experimental analysis.
These algorithms are experimentally faster than \GMeetPlus{}, but finding tight bounds for their runtime complexity is still an open problem.

\GMeetPlus{} is an enrichment of the simple abstract algorithm \GMeet{}~\cite{quintero-ramics-2020}, which is also the base for the alternative algorithms introduced in this paper and is presented here as Algorithm~\ref{alg:GMeet}.
The proof of correctness of \GMeet{} and the description of \GMeetPlus{} are found in the original paper~\cite{quintero-ramics-2020}.

Given an arbitrary lattice $\Lat$, let $\calF$ be the set of all functions defined on $\Lat$. \GMeet{} starts with the function $h \eqdef f \meetF g$, computed pointwise $h(a) = f(a) \meetp{\Lat} g(a)$,
which is not necessarily a join-endomorphism. Then, it iterates a loop that resolves conflicts, in whatever order they are found, until there are no conflicts at all. Recall that we refer to a conflict as a pair of elements $a,b \in \Lat$ not conforming the join-endomorphism property: $h(a \join b) = h(a) \join h(b)$.
The main invariants kept during the loop are that the function $h$ is an upper-bound of the target function $f\meetE g$, and that $h$ decreases strictly whenever a conflict is resolved.

\begin{compactFigure}[ht]
  \def\algorithmicindent{0.5em}
  \def\indent{\hspace{0.5em}}
  \noindent
  \begin{minipage}{0.46\textwidth}
  \begin{algorithm}[H]
    \caption{$\GMeet{}(h)$, $h\in\calF$.\\Particularly, $\GMeet{}(f \meetF g) = f\meetE g$.}\label{alg:GMeet}
    \footnotesize
    \begin{algorithmic}[1]
    \Procedure{GMeet}{$h$}
    \State {\bf While} $\exists a,b\in L$ with $h(a\join b) \neq h(a)\join h(b)$:
    \State\indent {\bf If} $h(a\join b) \sqgt h(a) \join h(b)$:
    \State\indent\indent $h(a\join b) \leftarrow h(a) \join h(b)$
    \State\indent {\bf Else}:
    \State\indent\indent $h(a) \leftarrow h(a) \meet h(a\join b)$
    \State\indent\indent $h(b) \leftarrow h(b) \meet h(a\join b)$
    \State \Return $h$ \Comment Maximal join-end. below the input
    \EndProcedure
    \end{algorithmic}
    \end{algorithm}
    \begin{algorithm}[H]
      \caption{$\GMeetMono{}(h)$, $h\in\calF$.}\label{alg:GMeetMono}
      \footnotesize
      \begin{algorithmic}[1]
      \Procedure{GMeetMono}{$h$}
    \State $h\gets $\MonoBelow{}($h$)
    \State {\bf While} $\exists a,b\in L$ with $h(a\join b) \sqgt h(a)\join h(b)$:
    \State\indent $c \leftarrow h(a) \join h(b)$
    \State\indent {\bf For} each $x \sqleq a\join b$:
    \State\indent\indent $h(x) \leftarrow h(x) \meet c$
    \State \Return $h$ \Comment Maximal join-end. below the input
      \EndProcedure
      \end{algorithmic}
      \end{algorithm}
  \end{minipage}
  \hfill
  \def\indent{\hspace{0.8em}}
  \noindent
  \begin{minipage}{0.46\textwidth}
    \begin{algorithm}[H]
    \caption{$\MonoBelow{}(h)$, $h\in\calF$.}\label{alg:monoBelow}
    \footnotesize
    \begin{algorithmic}[1]
    \Procedure{MonoBelow}{$h$}
    \State {\bf For} each $b \in L$, top-down order:
    \State\indent {\bf For} each children $a$ of $b$:
    \State\indent\indent $h(a) \gets h(a)\meet h(b)$
    \State \Return $h$ \Comment Maximal monotone below the input
    \EndProcedure
    \end{algorithmic}
    \end{algorithm}
  \begin{algorithm}[H]
    \caption{$\GMeetMonoLazy{}(h)$, $h\in\calF$.}\label{alg:GMeetMonoLazy}
    \footnotesize
    \begin{algorithmic}[1]
    \Procedure{GMeetMonoLazy}{$h$}
  \State $h\gets $\MonoBelow{}($h$)
  \State {\bf Do}:
  \State\indent $h_0 \gets h$
  \State\indent {\bf For} $a,b\in L$:
  \State\indent\indent $h(a\join b) \leftarrow h(a\join b)\meet (h(a) \join h(b))$
  \State\indent $h\gets $\MonoBelow{}($h$)
  \State {\bf While} $h\neq h_0$
  \State \Return $h$ \Comment Maximal join-end. below the input
    \EndProcedure
  \end{algorithmic}
  \end{algorithm}
  \end{minipage}
\end{compactFigure}

\GMeet{} was originally designed as an algorithm for computing $f\meetE g$, but it can serve for the more general purpose of finding the maximal join-endomorphism below a given arbitrary function $h\in\calF$.
This maximal join-endomorphism is always well defined as will be shown in Corollary~\ref{cor:maximalJoinBelow}, derived from Theorem~\ref{thm:maximalBelow}.

Algorithm~\ref{alg:GMeet} differs from the original version of \GMeet{} in that it takes a single function $h \in \calF$ as input instead of two $f,g \in \calF$.
This is done precisely to reflect the fact that \GMeet{} solves a more general problem, and the original proof of correctness of \GMeet{} suffices for proving the version presented here because said proof only uses $f$ and $g$ to set the starting point $f\meetF g$, and to define the target function $f\meetE g$, which coincides with the the maximal join-endomorphism below the starting point $f\meetF g$.

\begin{thm}\label{thm:maximalBelow}
Let $\calS \subseteq \calF$ be a sublattice of $\calF$ such that the join operator $\joinS$ in $\calS$ coincides with the pointwise join operator $\joinF$ in $\calF$.
For every $f\in \calF$, there is a unique maximal $h\in\calS$ with $h\sqleq f$.
\end{thm}
\begin{proof}
Suppose $h_1, h_2\in\calS$ are two different maximal functions in $\calS$ satisfying $h_1, h_2\sqleq f$, i.e. $h \eqdef h_1 \joinF h_2 \sqleq f$.
Since $\joinS = \joinF$ then $h\in\calS$, and since $h_1$ and $h_2$ are incomparable, then $h_1, h_2 \sqlt h \sqleq f$.
This contradicts that $h_1$ and $h_2$ were maximal on the first place.
\end{proof}

The following is an immediate result from the above theorem.

\begin{cor}\label{cor:maximalJoinBelow}
For any $f\in \calF$, there is a unique maximal $h\in\jhspace{\Lat}$ with $h\sqleq f$.
\end{cor}

Theorem~\ref{thm:maximalBelow} can also be used directly to find a maximal monotonic function below another given function because in the sublattice of monotonic functions, the join operator is the pointwise join.

\begin{cor}\label{cor:maximalMonoBelow}
For any $f\in \calF$, there is a unique maximal monotonic $h\in\calF$ with $h\sqleq f$.
\end{cor}

\MonoBelow{} (Algorithm~\ref{alg:monoBelow}) implements Corollary~\ref{cor:maximalMonoBelow} by computing the maximal monotonic function below a given one in $O(n+m)$, where $n = |\Lat|$ and $m$ is the number of (direct) child relations that exists between elements.
The algorithm assumes precomputation of the list of children for each element in the lattice, and a list in topological order, from top down to bottom.

\GMeetMono{} (Algorithm~\ref{alg:GMeetMono}) is an alternative algorithm to \GMeet{} that also implements Corollary~\ref{cor:maximalJoinBelow}.
It works by introducing an invariant to \GMeet{} that preserves the monotonicity of $h$ on each iteration of the main loop.
This is shown formally in Theorem~\ref{thm:GMeetMono}.

\begin{thm}\label{thm:GMeetMono}
  $\GMeetMono{}$ computes the unique maximal join-endomorphism below the input $h$.
\end{thm}
\begin{proof}
Let $h_0\in\calF$ be the input of the algorithm, and $h^*\in\jhspace{\Lat}$ the unique maximal join-endomorphism satisfying $h^*\sqleq h_0$, i.e. the target output.
The algorithm works with the invariant property that $h$ is monotonic and $h\sqgeq h^*$.
The first step that calls \MonoBelow{}, guarantees this invariant because, on the one hand, $h$ is monotonic, and on the other, since all join-endomorphisms are monotonic, the maximal monotonic function $h$ with $h\sqleq h_0$ satisfies $ h_0\sqgeq h \sqleq h^*$.

For analyzing the {\bf while} loop, let $h$ and $h'$ denote the function $h$ before and after an iteration.
Let us show that the invariant is preserved, that is, whenever $h$ is monotonic and $h\sqgeq h^*$, then $h'$ is monotonic and $h'\sqgeq h^*$.
Indeed, if there are $a,b\in \Lat$ with $h(a\join b) \sqgt h(a)\join h(b)$, then for all $x$ we have $h'(x) \eqdef h(x) \meet (h(a)\join h(b))$ whenever $x \cleq a\join b$ and $h'(x) \eqdef h(x)$ otherwise. In the first case,
\begin{align*}
h'(x) &= h(x) \meet (h(a)\join h(b))\\
&\sqgeq h^*(x) \meet (h^*(a)\join h^*(b))\\
&= h^*(x) \meet (h^*(a\join b)) = h^*(x),
\end{align*}
hence $h'(x)\sqgeq h^*(x)$. In the second case, $h'(x) = h(x) \sqgeq h^*(x)$. Thus $h'$ satisfies $h' \sqgeq h^*$.
Moreover, $h'$ can be expressed as the pointwise meet between $h$ and the function that maps all elements below $a\join b$ to $h(a) \join h(b)$ and all other elements to the top element. 
Since both functions are monotone, it follows that $h'$ is also monotone, thus the invariant is preserved.
Moreover, the loop guarantees that $h'\sqlt h$ because $h'(a\join b) \sqlt h(a\join b)$, hence, in addition to preserve the invariant, the main loop terminates.
Termination occurs when no elements $a,b\in L$ exist satisfying the loop condition.
Since $h$ is monotone, termination happens if and only if $h(a\join b) = h(a)\join h(b)$ for all $a,b\in L$.
\end{proof}

\GMeetMonoLazy{} (Algorithm~\ref{alg:GMeetMonoLazy}) is a lazy variant of \GMeet{} that delays the transformation of $h$ into a monotonic function after the iteration over all pairs $a,b\in L$.

\begin{thm}\label{thm:GMeetMonoLazy}
  $\GMeetMonoLazy{}$ computes the unique maximal join-endomorphism below the input $h$.
\end{thm}
\begin{proof}
As in the proof of Theorem~\ref{thm:GMeetMono}, let $h$ and $h'$ be the functions before and after the iteration of the {\bf do-while} loop.
Let also $g$ be the function $h$ after the {\bf for} loop is executed and before the algorithm \MonoBelow{} is called, so that $h'=\MonoBelow{}(g)$.
Since \MonoBelow{} is called before each iteration, $h$ and $h'$ are always monotone functions.
To show that $h'\sqgeq h^*$, it suffices to show that $g\sqgeq h^*$ because $\jhspace{\Lat}$ is a sublattice of the lattice of monotone functions.
Moreover, by induction, letting $f$ and $f'$ be the function $h$ before and after each iteration of the {\bf for} loop, it suffices to show that whenever $f\sqgeq h^*$ then $f'\sqgeq h^*$.
This holds because
\begin{align*}
f'(a\join b) &= f(a\join b) \meet (f(a)\join f(b))\\
&\sqgeq f(a\join b) \meet (h^*(a)\join h^*(b))\\
&= f(a\join b) \meet (h^*(a\join b))\\
&\sqgeq f(a\join b) \meet f(a\join b).
\end{align*}
Thus all $f', g$ and $h'$ are upper bounds of $h^*$.
Termination occurs when $h'=h$, which happens if and only if $h=g=h'$, if and only if $h(a\join b) = h(a)\join h(b)$ for all $a,b\in L$. 
\end{proof}

The main contribution of \GMeetMono{} and \GMeetMonoLazy{} over the existing algorithm \GMeetPlus{} is the empirical speed superiority.
Finding tight upper bounds for these two algorithms is not done in this paper and remains as an open theoretical problem.
A secondary contribution of the algorithms is that they approach the problem from a different theoretical perspective, which may lead to ideas for future faster algorithms.

\subsubsection{Experimental Results}
The runtime complexity of \GMeetMono{} and \GMeetMonoLazy{} has an upper bound of $O(n^4)$ because the number of updates per element can never exceed the number of elements in the lattice, but experimentally this bound seems to be very loose, with the real bound lying between $O(n^3)$ and $O(n^2)$, in fact, closer to the latter.

\begin{figure}[ht]
\includegraphics[width=0.95\textwidth]{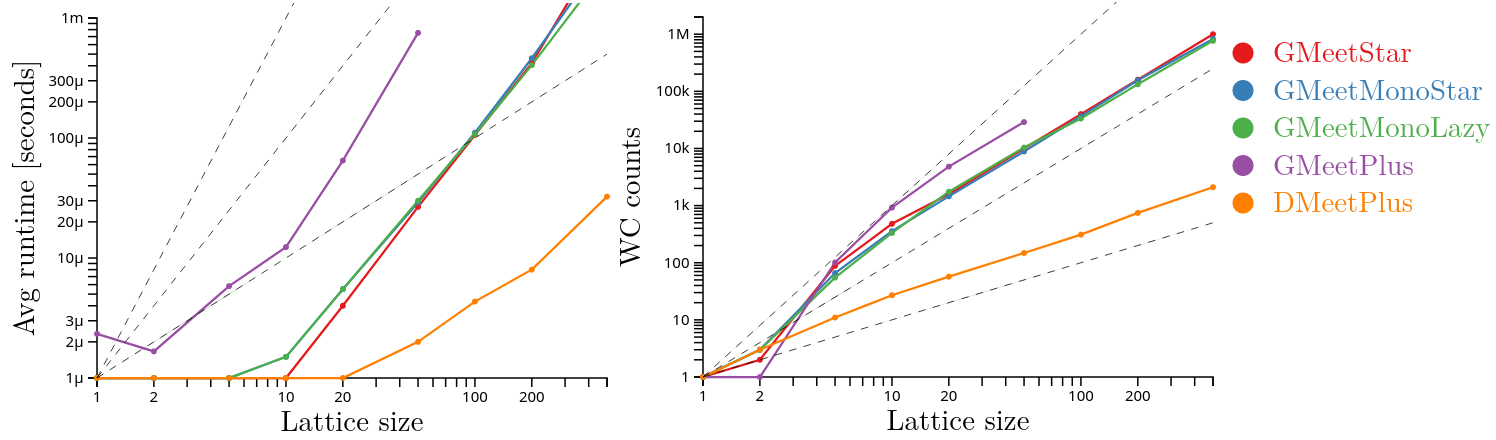}
\caption{Speed and worst case profiling counters for the algorithms of interest. Each point represents at least 6000 executions of each algorithm, varying the input functions and the lattices.}\label{fig:mono-experiments}
\end{figure}

Figure~\ref{fig:mono-experiments} shows the time and profiling counters for the algorithms on several experiments, as well as the three reference complexities $O(n)$, $O(n^2)$ and $O(n^3)$, which are straight lines in the log-log plane.
The counter units correspond to least upper bound and greatest lower bound calls.
Contrasting the slopes of the curves of the algorithms against the reference dashed lines and noting that any parallel line to one of them has the same complexity, the experiments suggest a behavior between $O(n^2)$ and $O(n^3)$, much closer to $O(n^2)$, for \GMeetStar{}, \GMeetMonoStar{} and \GMeetMonoLazy{}, which have very similar speeds in general.
The slope of average runtime of \GMeetPlus{} is more similar to $O(n^3)$ than $O(n^2)$, however, in the plot of worst case counters, it appears significantly smaller.
This difference is related with the fact that in addition to lattice operations, \GMeetPlus{} uses set operations extensively, which contribute to the runtime but not to the counters. 
\DMeetPlus{}, which is $O(n)$, is shown merely as a reference, since it yields incorrect outputs for non-distributive lattices.
The algorithms \GMeetStar{} and \GMeetMonoStar{} correspond to implementations of \GMeet{} and \GMeetMono{} respectively with a simple heuristic for executing the existential quantifier, namely, not restarting the search for $a,b\in L$ after each modification.
The implementations, as well as an interactive interface, are available at \url{https://caph1993.github.io/GMeetMono/}.

In this section we have presented algorithms to compute $f \meetp{\jhspace{\Lat}} g$ in distributive and arbitrary lattices. In doing so, we have exploited properties from lattice theory that allowed us to devise efficient procedures. Furthermore, we have provided a mechanism to find the maximal join-endomorphism below a given arbitrary function $h\in\calF$.
In the following section we will study some representation results for the latice of all join-endomorphisms defined on a lattice.

\section{A Representation of Join-Irreducible Elements of $\jhspace{\Lat}$}
\label{sec:j-endo-irred}

We now state a characterization of the join-irreducible elements of the lattice of join-endomorphisms $\jhspace{\Lat}$. We use it to prove a representation result for join-endomorphisms.
Some of these results may be part of  the folklore in lattice theory, our purpose here is to identify and use them as technical tools in the following section.

The following family of functions can be used to represent $\ijoin{\jhspace{\Lat}}$.

\begin{defi}
\label{def:jifuns}
Let $\Lat$ be a lattice and  $a,b \in \ijoin{\Lat}$. Let $\jifun{a,b} : \Lat \to \Lat$  be given by
$\jifunapp{a,b}{x} \defsymbol b$ if $x \in \upset{a}$, otherwise $\jifunapp{a,b}{x} \defsymbol \bote$.
\end{defi}

It is easy to verify that $\jifunapp{a,b}{\bote} = \bote$. On the other hand, for every $c,d \in \Lat$, $\jifunapp{a,b}{c \join d} = \jifunapp{a,b}{c} \join \jifunapp{a,b}{d}$ follows from the fact that $a \in \ijoin{\Lat}$ and by cases on $c \join d \in \upset{a}$ and $c \join d \not\in \upset{a}$.
Thus, from~\ref{prop:jendo-proprts1} we know
that $\jifun{a,b}$ is a join-endomorphism, and from~\ref{prop:jendo-proprts2} it is monotone. Therefore, $\fres{\jifun{a,b}}{\ijoin{\Lat}} \in \imonspace{\Lat}$.
In addition, we point out the following rather technical lemma that gives us way to construct from a function $g \in \imonspace{\Lat}$, a function $h \in  \imonspace{\Lat}$ covered by $g$.

\begin{lem}
\label{lemma:covers-imon}

Let $\Lat$ be a finite lattice. Let $g \in \imonspace{\Lat}$, $x_0 \in \ijoin{\Lat}$
and $y_0 \in \Lat$ be such that $y_0 \in \covers{g(x_0)}$ and
$g(x) \cleq y_0$ for all $x \cl x_0$.
Define $h:\ijoin{\Lat} \to \Lat$ as $h(x) \defsymbol y_0$ if $x = x_0$ else $h(x) \defsymbol g(x)$. 
Then $h$ is monotonic and $g$ covers $h$.
\end{lem}

\begin{proof}
    For notational convenience let $\fmvar=\imonspace{\Lat}$.
    We will prove (1) $h \in \fmvar$ and (2) $h \in \covers{g}$ in $\fmvar$.

    To prove (1), let $x_1,x_2 \in \ijoin{\Lat}$ with $x_1 \cl x_2$. We will show that $h(x_1) \cleq h(x_2)$.
    \begin{itemize}
    \item If both $x_1 \neq x_0$ and $x_2 \neq x_0$, then $h(x_1) = g(x_1) \cleq g(x_2) = h(x_2)$.
    \item If $x_0 = x_1 \cl x_2$, then $h(x_1) = y_0 \cl g(x_1) \cleq g(x_2) = h(x_2)$.
    \item If $x_1 \cl x_2 = x_0$, then $h(x_1) = g(x_1) \cleq y_0 = h(x_2)$.
    \end{itemize}
    
    Now we prove (2).
    From the definition of $h$, it follows that $h \mcl g$. If there is a function $\bar{h} \in \fmvar$ such that $h \mcl \bar{h} \mcl g$, then it must be the case that $\bar{h}(x) = g(x)$ for all $x \in \ijoin{\Lat}$ with $x \neq x_0$ and $h(x_0) \cl \bar{h}(x_0) \cl g(x_0)$, which is impossible since $h(x_0) = y_0$ and $y_0 \in \covers{g(x_0)}$.
    
    Thus, we conclude $g$ covers $h$ in $\fmvar$.
\end{proof}

We proceed to characterize the join-irreducible elements of the lattice $\jhspace{\Lat}$. The next lemma, together with~\ref{prop:jendo-proprts5}, tell us that every join-endomorphism in ${\jhspace{\Lat}}$  can be expressed solely as a join of functions 
of the form $f_{a,b}$ defined in Definition~\ref{def:jifuns}.

\begin{lem}
\label{lem:j-endo-irred}

Let $\Lat$ be a finite distributive lattice. For any join-endomorphism $f \in \jhspace{\Lat}$,
$f$ is join-irreducible 
iff $f = \jifun{a,b}$  for some $a,b \in \ijoin{\Lat}$.
\end{lem}

\begin{proof} For notational convenience let $\fmvar=\imonspace{\Lat}.$ 
From~\ref{prop:jendo-proprts4} it suffices to prove:
$g \in \fmvar$ is join-irreducible in $\fmvar$ iff $g =
g_{a,b}$ for some $a,b \in \ijoin{\Lat}$ where $g_{a,b} =
\fres{\jifun{a,b}}{\ijoin{\Lat}}$. We use the following immediate consequence of Lemma~\ref{lemma:covers-imon}.

\textbf{Property} $(\star)$: Let $g \in \fmvar$, $x_1,x_2 \in \ijoin{\Lat}$
and $y_1,y_2 \in \Lat$, be such that for each $i \in \{1,2\}$,
$y_i \in \covers{g(x_i)}$ and $g(x) \cleq y_i$ for all $x \cl x_i$.
If $x_1 \neq x_2$ or $y_1 \neq y_2$, then there are two distinct 
functions $g_1, g_2 \in \fmvar$ that are covered by $g$ in $\fmvar$.

\begin{enumerate}
\item For the only-if direction, let $X = \{x \in \ijoin{\Lat} \mid g(x)
\neq \bote\}$ and $Y = \{g(x) \mid x \in X\}$. If $X = \emptyset$, then
$g(x) = \bote$ for all $x \in \ijoin{\Lat}$, in which case $g$ is not 
join-irreducible in $\fmvar$. Thus, necessarily, $X \neq \emptyset$
and $Y \neq \emptyset$.
Let us now prove that:
(a) $X$ has a minimum element $a \in \ijoin{\Lat}$ with $g(a) \in \ijoin{\Lat}$, and
(b) $Y = \{g(a)\}$.
\begin{enumerate}

\item 
Let $x_1,x_2 \in X$ be minimal elements in $X$. For each $i \in \{1,2\}$, let
$y_i \in \covers{g(x_i)}$. Since $x_i$ is minimal, it follows that
$g(x) = \bote$ for all $x \cl x_i$. From ($\star$) and the fact that $g$ is
join-irreducible, we have $x_1 = x_2$ and $y_1 = y_2$.
Thus, $X$ has a minimum element. We refer to such element as $a$.
Furthermore, $|\covers{g(a)}|=1$, i.e. $g(a) \in \ijoin{\Lat}$.

\item 
Let $Y^* = Y \setminus \{g(a)\}$. For the sake of contradiction, suppose $Y^* \neq
\emptyset$. Let $y \in Y^*$ be a minimal element and $x^* \in X$ be a minimal of
$X^* = \{x \in X \mid g(x) = y\}$. Since $a \cl x^*$ and $y \neq g(a)$, we have
$g(a) \cl g(x^*) = y$. Then there is at least one $z \in \covers{y}$
such that $g(a) \cleq z \cl y$. Since $g$ is monotonic, $\im(g) = \{\bote\}
\cup Y$ and $y$ is minimal in $Y^*$, for all $x \cl x^*$, we have $g(x) \in
\{\bote,g(a)\}$. Therefore, $g(x) \cleq z$ for all $x \cl x^*$.
From ($\star$), with $x_1 = a$, $x_2 = x^*$, $y_1 \in \covers{g(a)}$ and
$y_2 = z$, it follows that $g$ is not join-irreducible in
$\fmvar$, a contradiction.
\end{enumerate}

Monotonicity of $g$ and (a)-(b), imply $\im(g) = \{\bote,b\}$
with $b = g(a)$. Thus $g = g_{a,b}$.

\item We prove that $g = g_{a,b}$ has a unique cover
in $\fmvar$. Let $c$ be the only cover of $b$.
Define $g^* {:} \ijoin{\Lat} \to \Lat$ as $g^*(x) = c$ if $x = a$ else
$g^*(x) = g(x)$. From Lemma~\ref{lemma:covers-imon}, it follows
that $g^* \in \fmvar$ and $g_{a,b}$ covers $g^*$ in $\fmvar$. 
It suffices that for any $h \in \fmvar$ with $h \mcl g_{a,b}$, $h \mleq g^*$ holds. Take any such $h \in \fmvar$.
Since $h(a) \neq b$, $h(a) \cl b$.
Thus $h(a) \cleq c$, so $h(a) \cleq g^*(a).$ Indeed,  
for any $x \neq a$, $h(x) \cl g(x) = g^*(x)$. Then $h \mleq g^*$.\qedhere
\end{enumerate}
\end{proof}

We conclude with a corollary of Lemma~\ref{lem:j-endo-irred} that provides a representation theorem for join-endomorphism on distributive lattices. 
We will use this result in the next section.

\begin{cor}
\label{cor:char-j-endo}
Let $\Lat$ be a finite distributive lattice and let 
$f \in \jhspace{\Lat}$. Then $f = F_R$ where 
$R =  \{(a,b) \in \ijoin{\Lat}^2\mid a \cleq f(b)\}$
and $F_R : \Lat \to \Lat$ is the function given by $F_R(c) \defsymbol \bigjoin \{a \in \ijoin{\Lat} \mid 
(a,b) \in R \text{ and } c \cgeq b \text{ for some } b \in \ijoin{\Lat}\}$.
\end{cor}

\begin{proof}
From~\ref{prop:jendo-proprts5}, we have
$f = \bigjoinp{\jhspace{\Lat}} \left\{g \in \ijoin{\jhspace{\Lat}}\mid g \hleq f \right\}$. Thus,
\begin{align*}
f(c) =&\ \left( \bigjoinp{\jhspace{\Lat}} \left\{g \in \ijoin{\jhspace{\Lat}}
\mid g \hleq f \right\} \right)(c)\\
=&\ \bigjoin \left\{g(c) \mid g \in \ijoin{\jhspace{\Lat}} \text{ and }
g \hleq f \right\} \\
=&\ \bigjoin \left\{\jifunapp{b,a}{c} \mid (b,a) \in \ijoin{\Lat}^2
\text{ and } \jifun{b,a} \hleq f \right\}\tag{Lemma~\ref{lem:j-endo-irred} }\\
=&\ \bigjoin \left\{\jifunapp{b,a}{c} \mid (b,a) \in \ijoin{\Lat}^2
\text{ and } a \cleq f(b) \right\}\\
=&\ \bigjoin \left\{a \in \ijoin{\Lat} \mid (b,a) \in \ijoin{\Lat}^2, a \cleq f(b) \text{ and } c \cgeq b \text{ for some }
b \in \ijoin{\Lat}\right\}\\
=&\ \bigjoin \left\{a \in \ijoin{\Lat} \mid (b,a) \in R \text{ and }
c \cgeq b \text{ for some } b \in \ijoin{\Lat}\right\} =  F_R(c)\qedhere
\end{align*}
\end{proof}

To summarize, we have provided a characterization of the join-irreducible elements of the lattice $\jhspace{\Lat}$ when $\Lat$ is distributive. Moreover, we have established a representation of join-endomorphisms in terms of a binary relation defined on the set of join-irreducible elements of $\Lat$ that satisfy some concrete conditions. The next section will study the relationship between join-endomorphisms and operators that represent knowledge.

\section{Distributive Lattices and Knowledge Structures}
\label{sec:app-know}

In this section, we introduce some knowledge structures from economics~\cite{aumann-disagree-1976,samet-agreeing-2010} and relate them to distributive lattices by adapting fundamental duality results between modal algebras and frames~\cite{bao-tarskip2-1952}.  
We will use these  structures and their relation to distributive lattices in the algorithmic results in the next section. 
We use the term \emph{knowledge} to encompass various epistemic concepts including $S5$ knowledge and belief~\cite{fagin1995reasoning}. 

\begin{defi}[\cite{samet-agreeing-2010}]
A (finite) \emph{Knowledge Structure (KS)} for a set of \emph{agents} $\Aindex$ is a tuple $(\Stateset, \{ \Kfun{i} \}_{i \in \Aindex})$ where $\Stateset$ is a finite set  and each $\Kfun{i}:\pset{\Omega}\to\pset{\Omega}$ is given by $\Kfunapp{i}{E} = \{ \state \in \Stateset \ | \ \Relapp{i}{\state} \subseteq E \}$ where $\Rel{i} \subseteq \Stateset^2$  and $\Relapp{i}{\state} = \{ \state' \ | \ (\state,\state')\in\Rel{i}\}.$ 
\end{defi}

The elements $\state \in \Stateset$ and the subsets $E \subseteq \Stateset$ are called \emph{states} and \emph{events}, resp. We refer to $\Kfun{i}$ and $\Rel{i}$ as the \emph{knowledge operator} and the \emph{accessibility relation} of agent $i$.

The notion of \emph{event} may be familiar to some readers from probability theory; for example the event \emph{``public transportation is suspended"} corresponds the set of states at which public transportation is suspended.  An event $E$ \emph{holds at} $\state$ if $\state \in E$.
Thus $\Stateset$, the event that holds at every $\state$,  corresponds to true in logic,  union of events corresponds to disjunction, intersection to conjunction, and complementation in $\Stateset$ to negation. We use  $\overline{E}$ for $\Stateset \setminus E$. We write $E \Eimply F$ for the event  $\overline{E} \cup F$ which corresponds to classic logic implication. We say that $E$ \emph{entails }$F$ if $E\subseteq F$. The event of \emph{$i$ knowing $E$} is $\Kfunapp{i}{E}.$

The following properties hold for all events $E$ and $F$ of any KS $(\Stateset, \{ \Kfun{i} \}_{i \in \Aindex})$:

\begin{enumerate}[start=1,label={($\kfrk$\arabic*)}]
  \item $\Kfunapp{i}{\Stateset}=\Stateset$\label{kp1},
  \item $\Kfunapp{i}{E} \cap \Kfunapp{i}{F} = \Kfunapp{i}{E \cap F}$\label{kp4},
  \item $\left( \Kfunapp{i}{E} \cap \Kfunapp{i}{E \Eimply F} \right) \subseteq \Kfunapp{i}{F}$\label{kp2}, and
  \item if $E \subseteq F$ then $\Kfunapp{i}{E} \subseteq \Kfunapp{i}{F}$\label{kp3}.
\end{enumerate}

Property~\refkp{kp1} represents that agents know the event that holds at every state, namely $\Stateset$. A distinctive property of knowledge is~\refkp{kp4}, i.e. if an agent knows two events, she knows their conjunction. In fact, \refkp{kp4} implies~\refkp{kp2}, that expresses modus ponens for knowledge. Other property implied by \refkp{kp4} is \refkp{kp3}, meaning that knowledge is monotonic, i.e. agents know the consequences of their knowledge.

An agent $i$ is \emph{wiser} (or \emph{more knowledgeable}) than $j$ iff $\Kfunapp{j}{E} \subseteq \Kfunapp{i}{E}$ for every event $E$; i.e. if $j$ knows $E$ so does $i$.

{\bf Aumann Structures.} 
Aumann structures are the standard event-based formalism in economics and decision theory~\cite{fagin1995reasoning} for reasoning about knowledge. \emph{A (finite) {Aumann structure (AS)} is a KS  where all the accessibility relations are equivalences.}\footnote{The presentation of AS~\cite{aumann-disagree-1976} uses a partition $\Par{i} = \{ \Relapp{i}{\state} \ | \ \state \in \Stateset \}$ of $\Stateset$ and  $\Kfunapp{i}{E}$ is  equivalently defined as $\{ \state \in \Stateset \ | \ \Parapp{i}{\state} \subseteq E \}$ where  $\Parapp{i}{\state}$ is the cell of $\Par{i}$ containing $\state$.}
The intended notion of knowledge of AS is $S5$; i.e. the knowledge captured by properties \refkp{kp1}-\refkp{kp4} and the following  three fundamental properties which hold for any AS:
\begin{enumerate}[start=5,label={($\kfrk$\arabic*)}]
  \item $\Kfunapp{i}{E}\subseteq E$\label{kp7},
  \item $\Kfunapp{i}{E}\subseteq\Kfunapp{i}{\Kfunapp{i}{E}}$\label{kp8}, and
  \item $\overline{\Kfunapp{i}{E}} \subseteq \Kfunapp{i}{\overline{\Kfunapp{i}{E}}}$\label{kp9}.
\end{enumerate}
The first says that if an agents knows $E$, then $E$ cannot be false; the second and third state that agents know both what they know and what they do not know. 

A straightforward property between knowledge operators and accessibility relations is that they uniquely define each other.

\begin{prop}
\label{prop:completeness}
Let $(\Stateset, \{ \Kfun{i} \}_{i \in \Aindex})$ be a KS and $i,j \in \Aindex$.
Then $\Kfun{i}=\Kfun{j}$  iff $\Rel{i}=\Rel{j}$. 
\end{prop}  

\begin{proof}
The ``if'' direction is obvious. For the other direction
suppose $\Kfun{i}=\Kfun{j}$ but $\Rel{i}\neq\Rel{j}$. Then there exists $\state$ such that $\Relapp{i}{\state}\neq\Relapp{j}{\state}$. If $\Relapp{i}{\state}$ is not included in $\Relapp{j}{\state}$  then we obtain $\state \not\in \Kfunapp{j}{\Relapp{i}{\state}}$ but $\state \in \Kfunapp{i}{\Relapp{i}{\state}}$, a contradiction with $\Kfun{i}=\Kfun{j}$. The case when $\Relapp{j}{\state}$ is not included in $\Relapp{i}{\state}$ is symmetric. 
\end{proof}

{\bf Extended KS.}  
We now introduce a simple extension of KS that will allow us to give a uniform presentation of our results. 
\begin{defi}[EKS]
\label{eks:def}
A tuple $(\Stateset, \scal, \{ \Kfun{i} \}_{i \in \Aindex})$ is said to be an \emph{extended knowledge structure (EKS)}  if (1)  $(\Stateset, \{ \Kfun{i} \}_{i \in \Aindex})$ is a KS, and (2)
$\scal$ is a  subset  of $\pset{\Stateset}$  that contains $\Stateset$   and it is closed under union, intersection and application of $\Kfun{i}$ for every $i \in \Aindex$. 
\end{defi}
\emph{Notation.} Given an underlying EKS $(\Stateset, \scal, \{ \Kfun{i} \}_{i \in \Aindex})$ and $f:\pset{\Stateset}\to\pset{\Stateset}$ we shall use $\Rfun{f}$ for the function $\fres{f}{\scal}:\scal 
\to \pset{\Stateset}$,  i.e.  $\Rfun{f}(E) = f(E)$ for every $E \in \scal$. Because of
the closure properties of  $\scal$,  for every $i\in\Aindex$ we have $\RKfun{i}:\scal \to \scal.$  
  
Notice that the AS and, in general KS, are EKS where $\scal = \pset{\Stateset}$. Also Kripke frames~\cite{fagin1995reasoning} can be viewed as EKS with $\scal = \pset{\Stateset}$. Other structures not discussed in this paper such as  set algebras with operators (SOS)~\cite{samet-s5knowledge-2010} and general frames~\cite{chagrov-modal-1997} can be represented as EKSs where $\scal$ is required to be closed under complement. 

\subsection{Extended KS and Distributive Lattices}
\label{ssec:ks-distlat}
 
The knowledge operators of an EKS are join-endomorphisms on a distributive lattice. This is an easy consequence
of  \refkp{kp1} and \refkp{kp4}, and the closure properties of EKS.  The next proposition tells us that  the \emph{wiser} the agent, the lower that (its knowledge operator) is placed in the corresponding lattice.

\begin{prop}
\label{prop:eks-to-lat}
Let $(\Stateset, \scal, \{ \Kfun{i} \}_{i \in \Aindex})$ be an EKS. Then $\Lat = (\scal, \supseteq)$ is a distributive lattice and for each $i\in\Aindex$, $\RKfun{i} \in \jhspace{\Lat}$.
\end{prop}

\begin{proof}
Since $\scal$ is closed under union and intersection and, $\Stateset \in \scal$, $\Lat = (\scal, \supseteq)$ is a distributive lattice whose join is the intersection and  bottom is $\Stateset$. By definition $\RKfunapp{i}{E}=\Kfunapp{i}{E}$ for every $E \in \scal$.
Thus, from \refkp{kp1} and \refkp{kp4}, $\RKfun{i}(\Stateset)=\Stateset$ and $\RKfunapp{i}{E \cap F}= \RKfunapp{i}{E} \cap \RKfunapp{i}{F}$ for every $E,F \in \scal$. From Property~\ref{prop:jendo-proprts1}, we conclude $\RKfun{i} \in \jhspace{\Lat}$.
\end{proof}

Conversely, the join-endomorphisms of  distributive lattices correspond to knowledge operators of EKS.  Recall that every distributive lattice is isomorphic to (the dual of) a lattice of sets.
The next proposition is an adaptation to finite distributive lattices of J\'onsson-Tarski duality for general-frames and boolean algebras with operators~\cite{bao-tarskip2-1952}.   

\begin{prop}
\label{prop:lat-to-eks}
Let $\Lat$ be dual to a finite lattice of sets with a family $\{ f_i \in \jhspace{\Lat} \}_{i \in I}$. Then $(\Stateset, \scal, \{ \Kfun{i} \}_{i \in I})$  is an EKS where $\scal = \Lat, \Stateset=\bot_\Lat$, and for every $i\in I$,  $\Rel{i}=\{ (\state,\state') \in \Stateset^2  \ |  \ \mbox{ for all }E\in \scal, \state \in f_i(E) \mbox{ implies } \state' \in E \}$.
Furthermore, for every $i \in I$, $\RKfun{i} = f_i$.
\end{prop}

\begin{proof}
Notice that  $\Lat=\scal$ is closed under union and intersection since $\Lat$ is the dual of a lattice of sets. Showing $\RKfun{i} = f_i$ also proves that $\scal$ is closed under $\Kfun{i}$. 
Recall that $\RKfunapp{i}{E}=\Kfunapp{i}{E}$ for each $E \in \scal.$ Thus, it remains to prove $\Kfunapp{i}{E}=f_i(E)$ for all $E \in \scal.$  From  \refkp{kp1} and the fact that $f_i$ is a join-endomorphism,  $\Kfunapp{i}{E}=f_i(E)=\Stateset$ for $E=\Stateset.$ Hence, choose an arbitrary $E \neq \Stateset$.  
First suppose that $\tau \in f_i(E)$. From the definition of $\Rel{i}$ if $(\tau,\tau') \in \Rel{i}$, $\tau' \in E$. Hence $\Relapp{i}{\tau} \subseteq E$, so $\tau \in  \Kfunapp{i}{E}$.

Now suppose that $\tau \in \Kfunapp{i}{E}$ but $\tau \not\in f_i(E)$. From $\tau \in \Kfunapp{i}{E}$ we obtain:
\begin{equation}
\emph{for all $\tau' \in \Stateset$ if $(\tau,\tau') \in \Rel{i}$ then $\tau' \in E$.}\label{eq1}
\end{equation}
From the assumption $\tau \not\in f_i(E)$  and the monotonicity of join-endomorphisms (\ref{prop:jendo-proprts2}):
\begin{equation}
\emph{for every $F \in \scal$ if $F \subseteq E$ then $\tau \not\in f_i(F)$.}\label{eq2}
\end{equation}
Let $X = \{ E'  \in \scal \mid  \tau \in f_i(E')  \}.$ If $X=\emptyset$ then from the definition of $\Rel{i}$ we conclude $\Relapp{i}{\tau}=\Stateset$ which contradicts (\ref{eq1}) since $E \neq \Stateset$. If $X\neq \emptyset$ take $S = \bigcap X$.  
Since $f_i$ is a join-endomorphism, it distributes over intersection (i.e. the join in $\Lat$), we conclude $\tau \in f(S)$. Thus, if $S \subseteq E$ we obtain a contradiction with (\ref{eq2}). If $S \not\subseteq E$ then there exists $\tau' \in S$ such that $\tau' \not\in E$. From the definition of $S$, $\tau' \in E'$ for each $E'$ such that $\tau \in f_i(E')$. But this implies $(\tau,\tau') \in \Rel{i}$ and $\tau' \not\in E$, a contradiction with (\ref{eq1}).
\end{proof}

Nevertheless, we can use our general characterization of join endomorphisms in the previous section (Corollary~\ref{cor:char-j-endo}) to obtain a simpler relational construction for join endomorphisms of powerset lattices (boolean algebras).  Unlike the construction in Proposition~\ref{prop:lat-to-eks}, this characterization of $\Rel{i}$ does not appeal to universal quantification.  

\begin{prop}
\label{prop:lat-to-ks}
Let $\Lat$ be dual to a finite powerset lattice with a family $\{ f_i \in \jhspace{\Lat} \}_{i \in I}$. Let $(\Stateset, \{ \Kfun{i} \}_{i \in I})$ be the KS where $\Stateset = \bot_\Lat$ and  $\Rel{i}=\left\{(\state,\state') \ \mid \state \in \overline{f_i(\   \overline{\{\state'\}}  \ ) }\ \right\}$.   
Then, for every $i \in \Aindex$, $\Kfun{i} = f_i$.
\end{prop}

\begin{proof}
  Since $\Lat$ is dual to a powerset lattice, $\join = \cap$, $\cleq = \supseteq$, and $\ijoin{\Lat} = \left\{\overline{\{\tau\}} \,\big|\, \tau \in \Stateset \right\}$. Let $Q = \left\{(\overline{\{\sigma\}},\overline{\{\tau\}})\mid (\sigma,\tau) \in \Rel{i}\right\}$.
  Notice that for every $(\overline{\{\sigma\}},\overline{\{\tau\}}) \in Q$, we have $\sigma \in \overline{f_i(\  \overline{\{\tau\}} \ ) }$. Equivalently, $\{\sigma\} \subseteq \overline{f_i(\ \overline{\{\tau\}}  \ ) }$ and $f_i(\ \overline{\{\tau\}}  \ ) \subseteq \overline{\{\sigma\}}$. Therefore, from Corollary~\ref{cor:char-j-endo}, it follows that for every $E \in \Lat$,
  \[f_i(E) = \bigcap \left\{\overline{\{\sigma\}} \in \ijoin{\Lat} \mid (\overline{\{\sigma\}},\overline{\{\tau\}}) \in Q_i \text{ and } E \subseteq \overline{\{\tau\}} \text{ for some } \overline{\{\tau\}} \in \ijoin{\Lat}\right\}.\]
  We complete the proof as follows:
  \begin{align*}
  f_i(E) = &\ \bigcap \left\{\overline{\{\sigma\}} \in \ijoin{\Lat} \mid \exists \overline{\{\tau\}} \in \ijoin{\Lat}:
  ((\overline{\{\sigma\}},\overline{\{\tau\}}) \in Q \text{ and } E \subseteq \overline{\{\tau\}}) \right\}\\
  =&\ \bigcap \left\{\overline{\{\sigma\}} \in \ijoin{\Lat} \mid \neg\forall \overline{\{\tau\}} \in \ijoin{\Lat}:
  ((\overline{\{\sigma\}},\overline{\{\tau\}}) \in Q \implies E \not\subseteq \overline{\{\tau\}}) \right\}\\
  =&\ \bigcap \{\Stateset \setminus \{\sigma\} \in \ijoin{\Lat} \mid \neg\forall \tau \in
  \Stateset : ((\sigma,\tau) \in \Rel{i} \implies \tau \in E) \}\\
  =&\ \bigcap \{\Stateset \setminus \{\sigma\} \in \ijoin{\Lat} \mid
   \neg (\Rel{i}(\sigma) \subseteq E) \}\\
  =&\ \Stateset \setminus \{\sigma \in \Stateset \mid \neg (\Rel{i}(\sigma) \subseteq E) \}= \ \{\sigma \in \Stateset \mid \Relapp{i}{\sigma} \subseteq E \}
  = \Kfunapp{i}{E}\qedhere
\end{align*}
\end{proof}

We conclude this section by pointing out that accessibility relations can be obtained from knowledge operators. 

\begin{cor}
\label{cor:k-r}
Let $\kcal = (\Stateset, \{ \Kfun{i} \}_{i \in \Aindex})$ be a KS. Then
\begin{enumerate}
\item $\Rel{i}=\left\{(\state,\state') \ \mid \state \in \overline{\Kfunapp{i}{\ \overline{\{\state'\}}  \ } }\ \right\}$.
\item If $\kcal$ is an AS then $\Relapp{i}{\state} = \overline{\Kfunapp{i}{\ \overline{\{\state\}}  \ } }$ for every $\state \in \Stateset$.
\end{enumerate}
\end{cor}

\begin{proof}
The proof of (1) is an immediate consequence of Proposition~\ref{prop:completeness} and Proposition~\ref{prop:lat-to-ks}.
For (2) rewrite the property as $\Relapp{i}{\state'}= \overline{\Kfunapp{i}{\   \overline{\{\state'\}}  \ } }$ for every $\state' \in \Stateset.$
if $\kcal$ is an AS then $\Rel{i}$ is an equivalence. Thus from  the symmetry of $\Rel{i}$ and (1) we obtain:
$(\state',\state) \in \Rel{i}$ iff $(\state,\state') \in \Rel{i}$ iff $\state \in \overline{\Kfunapp{i}{\   \overline{\{\state'\}}}}$. This implies (2).
\end{proof}

In this section we have related knowledge structures and distributive lattices via duality. Namely, we have provided results that represent knowledge operators ---that formalize agents' knowledge--- as join-endomorphisms. In the following section we formalize the distributed knowledge of a given group as the meet of the knowledge of its members, represented by join-endomorphisms.

\section{Distributed Knowledge.}\label{sec:AK}

The notion of \emph{distributed knowledge} represents the information that  two or more agents may have as a group but not necessarily individually.
Intuitively, it is what someone who knows what each agent, in a given group, knows.
As described in~\cite{fagin1995reasoning}, while common knowledge  can be viewed as what ``any fool'' knows, distributed knowledge can be viewed as what a ``wise man'' would know. 

Let $(\Stateset, \{ \Kfun{i} \}_{i \in \Aindex})$ be a KS and $i,j \in \Aindex$.  The  \emph{distributed knowledge} of  $i$ and $j$ is represented by  $\DKfun{i,j}:\pset{\Stateset} \to \pset{\Stateset} $ defined as $\DKfunapp{i,j}{E} = \{ \state \in \Stateset \ | \ \Relapp{i}{\state} \cap \Relapp{j}{\state}  \subseteq E \}$ where $\Rel{i}$ and $\Rel{j}$ are the accessibility relations for $i$ and $j$.

The following property captures the notion of distributed knowledge by relating group to individual knowledge:
\begin{enumerate}[start=8,label={($\kfrk$\arabic*)}]
\item $\left( \Kfunapp{i}{E} \cap \Kfunapp{j}{E \Eimply F} \right) \subseteq \DKfunapp{i,j}{F}$\label{kp5}
\end{enumerate}

It says that if one agents knows $E$ and the other knows that $E$ implies $F$,  together  they have the distributed knowledge of $F$ even if neither agent knew $F$. 

\begin{exa}
Let $E$ be the event ``Bob's \emph{boss is working from home}" and $F$ be the event ``\emph{public transportation is suspended}". Suppose that agent Alice knows that Bob's boss is working from home (i.e. $\Kfunapp{\emph{A}}{E}$), and that agent Bob knows that his boss works from home only when public transportation is suspended (i.e. $\Kfunapp{\emph{B}}{E \Eimply F}$).
Thus, if  they told each other what they knew, they would have distributed knowledge of $F$ (i.e. $\DKfunapp{\emph{A,B}}{F}$). Indeed, $\Kfunapp{\emph{A}}{E} \cap \Kfunapp{\emph{B}}{E \Eimply F}$ entails $\DKfunapp{\emph{A,B}}{F}$ from \refkp{kp5}.
\end{exa}

A self-explanatory property relating individual and distributed knowledge is \kproperty{ \Kfunapp{i}{E}  \subseteq \DKfunapp{i,j}{E}.}\label{kp6} 
Furthermore, the above basic properties of knowledge Proposition~\refkp{kp1}-\refkp{kp4}  also hold if we replace the $\Kfun{i}$ with $\DKfun{i,j}$:
Intuitively, distributed knowledge is knowledge. Indeed, imagine an agent $m$ that combines $i$ and $j$'s knowledge by having an accessibility relation $\Rel{m}=\Rel{i} \cap \Rel{j}.$ In this case we would have $\Kfun{m} = \DKfun{i,j}$. Therefore, any KS may include distributed knowledge as one of its knowledge operators. For simplicity, we are considering distributed knowledge of two agents but this can be easily extended to arbitrary groups of agents. E.g.  if $\Kfun{m} = \DKfun{i,j}$ then $\DKfun{k,m}$ represents the distributed knowledge of three agents $i,j$ and $k$. 

\subsection{The Meet of Knowledge.} In Section~\ref{ssec:ks-distlat} we identified knowledge operators and join endomorphisms. We now show that the notion of distributed knowledge corresponds exactly to the  meet of the knowledge operators in the lattice of all join-endomorphisms in $(\scal,  \supseteq )$.
\begin{thm}
\label{th:dk-as-meet}
Let $(\Stateset, \scal, \{ \Kfun{i} \}_{i \in \Aindex})$ be an EKS and let $\Lat$ be the lattice $(\scal,  \supseteq )$.  
Let us suppose that  $\Kfun{m}=\DKfun{i,j}$ for some $i,j,m\in\Aindex.$ Then  \( \RKfun{m} = \RKfun{i} \meetp{\jhspace{\Lat}} \RKfun{j}. \)
\end{thm}
\begin{proof}
Let us assume $\Kfun{m}=\DKfun{i,j}$. Then from the closure properties of $\scal$,  we have $\RDKfun{i,j}=\RKfun{m}:\scal \to \scal.$ 
Let $f =  \RKfun{i} \meetp{\jhspace{\Lat}} \RKfun{j}$.  (Recall that  the order relation $\cleq_\Lat$ over $\Lat$ is reversed inclusion $\supseteq$, joins are intersections and meets are unions.)

From Proposition~\refkp{kp6}, for every $E \in \scal$, $\DKfunapp{i,j}{E} \cleq_\Lat \Kfunapp{i}{E},\Kfunapp{j}{E}$. Thus $\RDKfun{i,j}$ is a lower bound of both $\RKfun{i}$ and $\RKfun{j}$ in ${\jhspace{\Lat}}$, so $\RDKfun{i,j}\cleq_{\jhspace{\Lat}}f.$

To prove $f \cleq_{\jhspace{\Lat}} \RDKfun{i,j}$, take $\tau \in \RDKfunapp{i,j}{E}=\DKfunapp{i,j}{E}$ for an arbitrary $E \in \scal$. 
By definition of $\DKfun{i,j}$, we have \inlineeq{\Relapp{i}{\tau} \cap \Relapp{j}{\tau} \subseteq E}\label{eq3}. From Proposition~\ref{prop:comp-algo}
 \begin{equation}\label{eq4}
f(E) = \bigcup\left\{ \Kfunapp{i}{F} \cap \Kfunapp{j}{H} \mid F,H \in \scal \mbox{ and } F \cap H \subseteq E \right\}
\end{equation}
Take $F= \Relapp{i}{\tau}$ and $H=\Relapp{j}{\tau}$, from (\ref{eq3}), $F \cap H \subseteq E$. By definition of knowledge operator, $\tau \in \Kfunapp{i}{F}$ and $\tau \in  \Kfunapp{j}{H}$. From (\ref{eq4}), $\tau \in f(E)$. Thus $f \cleq_{\jhspace{\Lat}} \RDKfun{i,j}$.
\end{proof}
The theorem above allows us to characterize an agent $m$ having the distributed knowledge of $i$ and $j$ as the \emph{least knowledgeable} agent wiser than both $i$ and $j$. In the next section we consider the decision problem of whether a given $m$ indeed has the distributed knowledge of $i$ and $j$.

\subsection{\bf The Distributed Knowledge Problem.} In what follows, let $(\Stateset, \{ \Kfun{i} \}_{i \in \Aindex})$ be a KS and let $n=|\Stateset|$. Let us now consider the following decision: \emph{Given the knowledge of agents $i,j,m$, decide whether $m$ has the distributed knowledge of $i$ and $j$, i.e.  $\Kfun{m}=\DKfun{i,j}$.}

The input for this problem is the knowledge of the agents and it can be represented using either knowledge operators $\Kfun{i},\Kfun{j},\Kfun{m}$ or accessibility relations $\Rel{i},\Rel{j},\Rel{m}$. 
For each representation, the algorithm that solves the problem $\Kfun{m}=\DKfun{i,j}$ can be implemented differently.
For the first representation, it follows from Theorem~\ref{th:dk-as-meet} that $\Kfun{m}=\DKfun{i,j}$ holds if and only if $\Kfun{m}=\Kfun{i}\meetp{\jhspace{\Lat}}\Kfun{j}$  where $\Lat=(\pset{\Stateset},\supseteq)$. 
For the second one, we can verify $\Rel{m}=\Rel{i}\cap \Rel{j}$ instead.
Indeed, as stated in Corollary~\ref{cor:k-r}, one representation can be obtained from the other, hence an alternative solution for the decision problem is to translate the input from the given representation into the other one before solving.

Accessibility relations represent knowledge much more compactly than knowledge operators because the former are relations on $\Stateset^2$ while the latter are relations on $\pset{\Stateset}^2$.
For this reason, it would seem in principle that the algorithm for handling the knowledge operator would be slower by several orders of magnitude.
Nevertheless, we can use our lattice theoretical results from previous sections to show that this is not necessarily the case,
thus it is worth considering both types of representations.

{\it From Knowledge Operators.} We wish to determine $\Kfun{m}=\DKfun{i,j}$ by establishing whether $\Kfun{m}=\Kfun{i}\meetp{\jhspace{\Lat}}\Kfun{j}$ where $\Lat=(\pset{\Stateset},\supseteq)$.  Let us assume the following bitwise representation of knowledge operators. The states in $\Stateset$ are numbered as  $\state_1,\ldots,\state_n$. Each event $E$ is represented as a number $\#E\in \Zinterval{0}{2^n-1}$  whose binary representation has its $k$-th bit set to 1 iff $\state_k \in E$. Each input knowledge operator $\Kfun{i}$ is represented as an array $\VKfun{i}$ of size $2^n$ that stores $\#\Kfunapp{i}{E}$ at position $\#E$, i.e. $\VKfunapp{i}{\ \#E\ }=\#\Kfunapp{i}{E}.$  

From Lemma~\ref{lem:jimeet}, $\Kfun{m}=\Kfun{i}\meetp{\jhspace{\Lat}}\Kfun{j}$ iff $\Kfunapp{m}{E}=\Kfunapp{i}{E} \cup \Kfunapp{j}{E}$ for every join-irreducible element $E$ in $\Lat$.  Notice that $E \in \ijoin{\Lat}$  iff $E$ has the form $\overline{\{\state_k\}}$ for some $\state_k \in \Stateset$. Moreover,  $\#\overline{\{\state_k\}}= (2^n - 1)-2^k$. These facts lead us to the following result.

\begin{thm}\label{th:knowledge-from-K}
Given the arrays $\VKfun{i},\VKfun{j},\VKfun{m}$ where $i,j,m\in I$,  there is an effective procedure  that can decide $\Kfun{m}=\DKfun{i,j}$  in time $O(n^2)$ where $n=|\Stateset|.$
\end{thm}
\begin{proof}
Let $\Lat=(\pset{\Stateset},\supseteq)$. We have $\Kfun{m}=\DKfun{i,j}$ iff $\Kfun{m}=\Kfun{i}\meetp{\jhspace{\Lat}}\Kfun{j}$ (Theorem~\ref{th:dk-as-meet}) iff $\Kfunapp{m}{E}=\Kfunapp{i}{E} \cup \Kfunapp{j}{E}$ for every $E \in \ijoin{\Lat}$ (Lemma~\ref{lem:jimeet}). Furthermore,  $E \in \ijoin{\Lat}$ iff $E = \overline{\{\state\}}$ for some $\state \in \Stateset$. Then we
can conclude that $E \in \ijoin{\Lat}$ iff $\#E = (2^n - 1)-2^k$ for some $k\in\Zinterval{0}{n-1}$. Therefore, $\Kfun{m}=\DKfun{i,j}$ iff for every 
$k\in \Zinterval{0}{n-1}$
\begin{equation}\label{eq5} \VKfunapp{m}{\ p_k\ } \ \ = \ \VKfunapp{i}{\ p_k \ } \  \ \ | \ \ \ \VKfunapp{j}{\ p_k\ } 
 \end{equation}
where $p_k=(2^n - 1)-2^k$ and $|$ is the OR operation over the bitwise representation of $\VKfunapp{i}{\ p_k \ }$ and $\VKfunapp{i}{\ p_k \ }$. For each $k\in \Zinterval{0}{n-1}$, the equality test and the OR operation in  Equation~\ref{eq5} can be computed
in $O(n)$. Hence the total cost is $O(n^2)$.
\end{proof} 

{\it From Accessibility Relations.}
A very natural encoding for accessibility relations is to use a binary $n\times n$ matrix.
If the input is encoded using three matrices $\Mtt_i,\Mtt_j$ and $\Mtt_m$, we can test whether $\Rel{m} = \Rel{i}\cap \Rel{j}$ (a proxy for $\Kfun{m}=\DKfun{i,j}$) in $O(n^2)$ by checking pointwise if $\Mtt_m[a,b] = \Mtt_i[a,b] \cdot \Mtt_j[a,b]$. 

It suggests that for \emph{AS} we can use a different encoding and check $\Rel{m} = \Rel{i}\cap \Rel{j}$ practically in \emph{linear time}: More precisely in  $O(\invack{n}n)$  where $\invack{n}$ is the inverse of the Ackermann function\footnote{Here $\invack{n}\eqdef\min\{k:A(k,k)\geq n\}$, where $A$ is the Ackermann function. The growth of $\invack{n}$ is negligible in practice, e.g. $\invack{n} = 4$ for $n=2^{2^{2^{65536}}}-3$.}. The key point is that the relations of AS are equivalences so they can be represented as \emph{partitions}. 

\subsection{An $O(n\invack{n})$ Algorithm for Partition Intersection}

The proof of the following result, which is interesting in its own right, shows an \(O(n\invack{n})\) procedure for deciding $\Rel{m} = \Rel{i}\cap \Rel{j}$.


\providecommand{\f}{}
\renewcommand{\f}[1][]{\mathtt{f}\arrindex{#1}}
\providecommand{\g}{}
\renewcommand{\g}[1][]{\mathtt{g}\arrindex{#1}}
\providecommand{\t}{}
\renewcommand{\t}[1][]{\mathtt{t}\arrindex{#1}}
\providecommand{\q}{}
\renewcommand{\q}[1][]{\mathtt{q}\arrindex{#1}}
\providecommand{\r}{}
\renewcommand{\r}[1][]{\mathtt{r}\arrindex{#1}}

\begin{thm}\label{th:knowledge-from-R}
  Let $\Rel{1}, \Rel{2}, \Rel{3} \subseteq \Stateset^2$ be equivalences over a set $\Stateset$ of $n=|\Stateset|$ elements.
  There is an $O(\invack{n} n)$ algorithm for the following problem:
  \begin{itemize}
    \item[] \textbf{Input:} Each $\Rel{i}$ in partition form, i.e. an array of disjoint arrays of elements of $\Stateset$, whose concatenation produces $\Stateset$. This is readable in $O(n)$.
    \item[] \textbf{Output:} Boolean answer to whether $\Rel{3} = \Rel{1}\cap\Rel{2}$.
  \end{itemize}
\end{thm}

\begin{proof}
We use the Disjoint-Sets data structure~\cite{galler1964improved}.
We can view a disjoint-set as a function $r:I\to I$ that satisfies $r\circ r = r$ and can be evaluated at a particular index in $O(\invack{n})$.
The element $r(i)$ corresponds to the class representative of $i$ for each $i\in I$, so that $i \sim_r j$ if and only if $r(i)=r(j)$.

If we let $r_i$ denote a disjoint-set for $\Rel{i}$ for each $i\in\{1,2,3\}$, and we let $q$ denote the disjoint-set for $\Rel{1}\cap\Rel{2}$, then the problem can be divided into computing the disjoint-set $q$ in $O(n\invack{n})$ and verifying whether $\sim_q = \sim_{r_3}$ also in $O(n\invack{n})$. To organize these claims, let us consider the following algorithm descriptions.

\begin{itemize}
  \item[] \textbf{Intersection.} Takes two disjoint-sets $r_1$ and $r_2$, and produces a disjoint-set $q$ such that $i \sim_q j$ iff $i\sim_{r_1} j$ and $i\sim_{r_2} j$.
  \item[] \textbf{Canonical.} Takes a disjoint-set $r$ and produces another $\hat r$ with $\sim_r=\sim_{\hat r}$, but such that $\hat{r}(i)\leq i$ for all $i\in I$.
  \item[] \textbf{Equality.} Takes two disjoint-sets $r_1,r_2$ and determines if $i\sim_{r_1} j$ iff $i\sim_{r_2} j$ for all $i,j\in I$. This problem is reduced simply to checking if $\hat{r}_1=\hat{r}_2$.
\end{itemize}

We proceed to show that Algorithms~\ref{alg:uf-intersection} and \ref{alg:uf-equality} compute $q$ and $\hat{r}$ (in array form) in $O(n\invack{n})$. The complexity follows from the fact that they must read the input function(s) pointwise and all other operations are linear. It remains to show correctness only.

\begin{compactFigure}[t]
\def\algorithmicindent{0.5em}
\noindent
\begin{minipage}{0.46\textwidth}
\begin{algorithm}[H]
  \caption{Intersection of disjoint sets in $O(n\invack{n})$}\label{alg:uf-intersection}
  \footnotesize
  \begin{algorithmic}[1]
    \Procedure{Intersection}{$r_1$, $r_2$}
    \State Let $\f:I\to I\times I$ be an array
    \State {\bf For} each $i\in I$ {\bf do} \\
    \hspace{\algorithmicindent}
    \hspace{\algorithmicindent}
    $\f[i] \gets (r_1(i),r_2(i))$
    \State Let $\g:\im(\f)\to I$ be a hash map
    \State {\bf For} each $i\in I$ {\bf do} $\g[\f[i]] \gets i$
    \State Let $\q:I\to I$ be an array
    \State {\bf For} each $i\in I$ {\bf do} $\q[i] \gets \g[\f[i]]$
    \State \Return $\q$
    \EndProcedure
  \end{algorithmic}
  \end{algorithm}
\end{minipage}
\hfill
\begin{minipage}{0.46\textwidth}
\begin{algorithm}[H]
  \caption{Equality of disjoint sets in $O(n\invack{n})$}\label{alg:uf-equality}
  \footnotesize
  \begin{algorithmic}[1]
    \Procedure{Canonical}{$r$}
    \State (Comment) $J\eqdef\{r(i):i\in I\}$.
    \State Let $\t:J\to I$ be a hash map.
    \State {\bf For} each $i\in I$ {\bf do} $t[r(i)] \gets r(i)$.
    \State {\bf For} each $i\in I$ {\bf do}\\
    \hspace{\algorithmicindent}
    \hspace{\algorithmicindent}
    $t[r(i)] \gets \min(t[r(i)], i)$
    \State Let $\hat{\r}:I\to I$ be an array
    \State {\bf For} each $i\in I$ {\bf do} $\hat{\r}[i] \gets t_{r(i)}$
    \State \Return $\hat{\r}$
    \EndProcedure
  \end{algorithmic}
\end{algorithm}
\end{minipage}
\end{compactFigure}

The array $\g$ in Algorithm~\ref{alg:uf-intersection} is any version of the inverse image of $\f$, i.e. $\f[\g[y]]=y$ for every $y\in\im(\f)$. This guarantees $\f \circ \g\circ \f = \f$ and hence 
\(
  \q\circ \q = \g\circ \f \circ \g\circ \f = \g\circ \f = \q
\).
Moreover, for any $i,j\in I$, $\q[i]=\q[j]$ iff $\g[\f[i]]=\g[\f[j]]$ by definition; iff $\f[i]=\f[j]$ because $\f$ is injective; iff $r_1(i)=r_1(j)$ and $r_2(i)=r_2(j)$; iff $i\sim_{r_1} j$ and $i\sim_{r_2} j$.

Regarding Algorithm~\ref{alg:uf-equality}, for all $i\in I$,  $i\sim t[r(i)]$, thus $r(i)=r(t[r(i)])$. This is, $r = r\circ \t\circ r$. Thus,
\(
  \hat{\r}\circ \hat{\r} = \t\circ r\circ \t\circ r = \t\circ r = \hat{\r}
\).
Moreover, for any $i,j\in I$, $i\sim j$ iff $r(i)=r(j)$; iff $t[r(i)]=t[r(j)]$ since $\t$ is injective on $J$; iff $\hat{\r}[i]=\hat{\r}[j]$ by definition.
\end{proof}

\subsubsection{Experimental Results.}

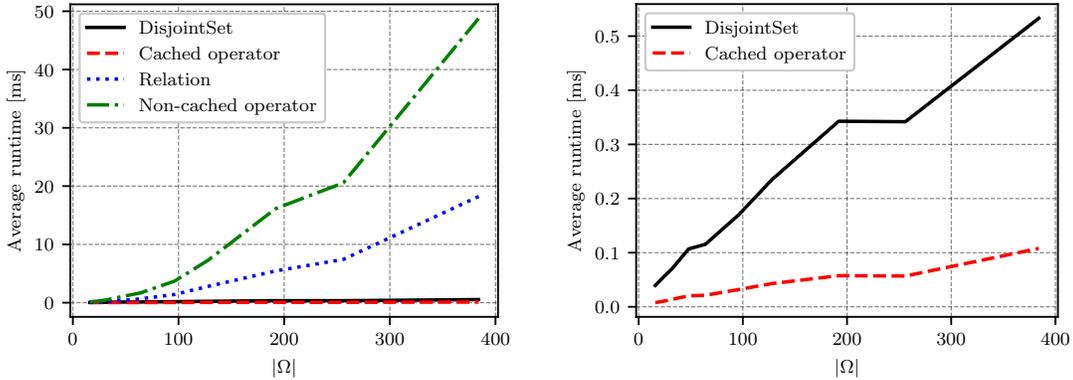
\begin{figure}[t]
    \centering
    \resizebox{0.48\textwidth}{!}{\input{figures/knowledge-cmp1.pgf}}
    \resizebox{0.48\textwidth}{!}{\input{figures/knowledge-cmp2.pgf}}
    \caption{
      Runtime comparison of several algorithms that solve the distributed knowledge problem.
    }
    \label{fig:knowledge-cmp}
  \end{figure}
  
Figure~\ref{fig:knowledge-cmp} shows the average runtime (100 random executions) of the four algorithms listed below for the distributed knowledge problem.
Fixing the number of elements $n=|\Stateset|$ elements, the input for each execution consisted of three randomly generated partitions $P_i$, $P_j$ and $P_m$.
The first two are generated independently and uniformly over the set of all possible partitions of $n$ elements.
The third, $P_m$, corresponds with $50\%$ probability to the intersection of the relations of the first two, and to a different but very similar partition otherwise, so as to increase the problem difficulty.
  
\begin{enumerate}
  \item The “Cached operator” algorithm is the one described in Theorem~\ref{th:knowledge-from-K}.
  It assumes that the input knowledge operators can be evaluated in $O(1)$ at any join-irreducible input $E\subseteq\Stateset$.
  Its complexity is $O(n^2)$, because bit-mask operations are linear w.r.t. the number of bits.
  However, this is compensated heavily in practice by the speed of bit-masking operations, at least for the sizes depicted.
  \item The “Disjoint set” algorithm is the one described in Theorem~\ref{th:knowledge-from-R} ($O(n  \invack{n})$).
  It takes the accessibility relations in partition form as input.
  \item The “Relation” algorithm ($O(n^2)$) takes as input the accessibility relations in the form of $n\times n$ binary matrices, and simply verifies if the pointwise-and matches.
  \item The “Non-cached operator” ($O(n^2)$) algorithm is that of the “Cached operator” when the cost of evaluating $\Kfun{i}(\,\cdot\,)$ is taken into account.
  It shows that although the “Cached operator” algorithm is very fast, its speed depends heavily on the assumption that the knowledge operators are pre-computed.
\end{enumerate}

\section{Concluding Remarks and Related Work.}
\label{sec:related}

We have used some standard tools from lattice theory to characterize the notion of distributed knowledge 
and provide efficient procedures to compute the meet of join-endomorphisms.  Furthermore, we provide an algorithm to compute
the intersection of \emph{partitions} of a set of size $n$ in $O(n\invack{n})$. As illustrated in the introduction, this algorithm may have applications for graph connected components and other domains where the notion of partition and intersection arise naturally.    
 
In~\cite{quintero-ramics-2020} we proposed algorithms to compute $f \meetp{\jhspace{\Lat}} g$ with time complexities $O(n^3)$ for arbitrary lattices and $O(n^2)$ for distributive lattices. Here we have improved the bounds to $O(n^2)$ (experimentally) for arbitrary lattices and $O(n)$ for distributive lattices. The authors in~\cite{HABIB1996391} gave a method of logarithmic time complexity (in the size of the lattice) for meet operations. Since $\jhspace{\Lat}$ is isomorphic to $\ocal({\ijoin{\Lat}\times \ijoin{\Lat}^{\it op}})$ for a distributive lattice $\Lat$, finding $f \meetp{\jhspace{\Lat}} g$ with their algorithm would be in $O(\log_2(2^{n^2})) = O(n^2)$ in contrast to our linear bound.  Furthermore, we would need 
a lattice isomorphic to $\jhspace{\Lat}$  to find $f \meetp{\jhspace{\Lat}} g$ using their algorithm. This lattice can be exponentially bigger than $\Lat$~\cite{quintero-ramics-2020} which is the input to our algorithm. We also provided experimental results illustrating the performance of our procedures. We followed the work in \cite{jipsen-genlattice-2015} for generating random distributive lattices.  
 
The finite representation results we used in Sections~\ref{sec:j-endo-irred} and~\ref{sec:app-know} to obtain our main results are adaptations from standard results from duality theory.  
J\'onsson and Tarski~\cite{bao-tarskip1-1951,bao-tarskip2-1952} originally presented an extension of boolean algebras with operators (BAO), called canonical extensions, provided with some representation theorems.
Roughly speaking, the representation theorems state that (1) every relation algebra is isomorphic to a complete and atomic relation algebra and (2) every boolean algebra with operators is isomorphic to a complex algebra that is complete and atomic.
The idea behind this result, as was presented later by Kripke in~\cite{kripke-modal-1959}, basically says that the operators  can be recovered from certain binary relations and vice versa. Another approach to this duality was given by Goldblatt~\cite{GOLDBLATT-varieties-1989} where it is stated that the variety of normal modal algebras coincides with the class of subalgebras defined on the class of all frames.
Canonical extensions have been useful for the development of duality and algebra. J\'onsson proved an important result for modal logic in~\cite{jonsson-canonicity-1994} and the authors of~\cite{gehrke-distlat-2004,gehrke-arblat-2001,dunn-posets-2005} have generalized canonical extensions for BAOs to distributive and arbitrary bounded lattices and posets. 

Distributed knowledge was introduced in \cite{HalpernM90} and various axiomatization and expressiveness for it have been provided, e.g. in \cite{HakliN07,AgotnesW17}.  In terms of computational complexity, the satisfiability  problem for  epistemic logic with distributed knowledge ($S5^{D}$) has been shown to be PSPACE-complete \cite{fagin1995reasoning}. Nevertheless, we are not aware of any lattice theoretical characterization of distributed knowledge nor algorithms to decide if an agent has the distributed knowledge of others.

\bibliographystyle{alphaurl}
\bibliography{biblio}

\end{document}

%% file: packages.tex
\usepackage{xcolor}
\usepackage{amsmath}
\usepackage{amssymb}
\usepackage{graphicx}
\usepackage{listings}
\usepackage{stmaryrd}
\usepackage{enumitem}
\usepackage{tikz}
\usetikzlibrary{trees,arrows,arrows.meta,shapes,decorations.pathmorphing,backgrounds,positioning,fit,petri}
\usepackage{algorithm}
\usepackage[noend]{algpseudocode}
\usepackage{url}
\usepackage[font=small]{caption}
\usepackage{subcaption}
\usepackage{booktabs}

\usepackage{hyperref}
\usepackage{mathtools}

\makeatletter
\newbox\dottedarrow@box
\setbox\dottedarrow@box\hbox
  {%
    \begin{tikzpicture}[>=stealth]
      \draw[blue,thick,dotted,->] (0,0) -- (1.5em,0);
    \end{tikzpicture}%
  }
\newcommand*\dottedarrow
  {\relax\ifmmode\expandafter\dottedarrow@m\else\expandafter\dottedarrow@t\fi}
\newcommand*\dottedarrow@t[1][1.5em]
  {\resizebox{#1}{!}{\raisebox{.5ex}{\usebox\dottedarrow@box}}}
\newcommand*\dottedarrow@m[1][]
  {%
    \if\relax\detokenize{#1}\relax
      \mathchoice
        {\dottedarrow@t}
        {\dottedarrow@t}
        {\dottedarrow@t[1.1em]}
        {\dottedarrow@t[0.9em]}%
    \else
      \dottedarrow@t[#1]%
    \fi
  }
\makeatother

%% file: macros.tex
\newcommand{\Fvar}{{\fcal}}   
\newcommand{\jhvar}{{\ecal}}  
\newcommand{\sfunspace}[1]{\jhvar(#1)}  
\newcommand{\jhspace}[1]{\sfunspace{#1}}  
\newcommand{\imonspace}[1]{\langle \ijoin{\Lat} \to #1 \rangle}  

\newcommand{\Mtt}{\mathtt{M}}

\newcommand{\dapprox}{h}
\newcommand{\jifunvar}{f}

\newcommand{\jifun}[1]{\jifunvar_{#1}}

\newcommand{\jifunapp}[2]{\jifun{#1} \left( {#2} \right)}

\newcommand{\fres}[2]{#1{\upharpoonright}_{{\scriptscriptstyle #2}}} 		
\newcommand{\im}{\textit{Im}}

\newcommand{\Rfun}[1]{\tilde{#1}}

\newcommand{\Rvar}{\rcal}
\newcommand{\Kvar}{\mathnormal{K}}
\newcommand{\RKvar}{\widetilde{\Kvar}}
\newcommand{\VKvar}{\mathtt{K}}
\newcommand{\DKvar}{\mathnormal{D}}
\newcommand{\RDKvar}{\widetilde{\DKvar}}
\newcommand{\Pvar}{\pcal}

\newcommand{\Rel}[1]{\Rvar_{{#1}}}
\newcommand{\Relapp}[2]{\Rel{#1}{( #2 )}}

\newcommand{\Par}[1]{\Pvar_{#1}}
\newcommand{\Parapp}[2]{\Par{#1}{( #2 )}}

\newcommand{\Kfun}[1]{\Kvar_{#1}}
\newcommand{\Kfunapp}[2]{\Kfun{#1}{( #2 )}}
\newcommand{\RKfun}[1]{\RKvar_{#1}}
\newcommand{\RKfunapp}[2]{\RKfun{#1}{( #2 )}}
\newcommand{\VKfun}[1]{\VKvar_{#1}}
\newcommand{\VKfunapp}[2]{\VKfun{#1}{[#2]}}

\newcommand{\DKfun}[1]{\DKvar_{\{#1\}}}
\newcommand{\DKfunapp}[2]{\DKfun{#1}{\left( #2 \right)}}
\newcommand{\RDKfun}[1]{\RDKvar_{\{#1\}}}
\newcommand{\RDKfunapp}[2]{\RDKfun{#1}{\left( #2 \right)}}

\newcommand{\Aindex}{\acal}

\newcommand{\Eimply}{\Rightarrow}

\newcommand{\Stateset}{\Omega}
\newcommand{\state}{\omega} 

\newcommand{\pset}[1]{\pcal(#1) }

\newcommand{\Lat}{L}   
\newcommand{\M}{\mbox{\bf M}}   
\newcommand{\C}{\Con}                   
\newcommand{\Con}{\Lat}					

\newcommand{\bote}{\bot}	   

\newcommand{\join}{\sqcup}						

\newcommand{\bigjoin}{\bigsqcup}						
\newcommand{\bigjoinp}[1]{\bigjoin\nolimits_{{\scriptscriptstyle #1}}}

\newcommand{\meet}{\sqcap}						
\newcommand{\meetp}[1]{\meet_{{\scriptscriptstyle #1}}}
\newcommand{\bigmeet}{\bigsqcap}						
\newcommand{\bigmeetp}[1]{\bigmeet\nolimits_{{\scriptscriptstyle #1}}}

\newcommand{\cleq}{\sqsubseteq}					
\newcommand{\cleqp}[1]{\cleq_{\scriptscriptstyle #1}}					
\newcommand{\cgeq}{\sqsupseteq}		
\newcommand{\cl}{\sqsubset}			
\newcommand{\cg}{\sqsupset}			
\newcommand{\cov}{\prec}    

\newcommand{\hleq}{\cleq_{\scriptscriptstyle \jhvar}}         
 
\newcommand{\fmvar}{\textit{M}}  		
\newcommand{\mleq}{\cleq_{\scriptscriptstyle \fmvar}}         

\newcommand{\mcl}{\cl_{\scriptscriptstyle \fmvar}}         

\newcommand{\downset}[1]{\left\downarrow #1\right.}

\newcommand{\upset}[1]{\left\uparrow #1\right.}
\newcommand{\varijoin}{\mathcal{J}}
\newcommand{\ijoin}[1]{\varijoin(#1)}
\newcommand{\jdown}[1]{\left\downarrow^{{\scriptscriptstyle\varijoin}} #1\right.}
\newcommand{\covers}[1]{{\downarrow^{1}}{{#1}}}

\newcommand{\sop}{\ominus}

\def\DMeet{\texttt{DMeet}}
\def\DMeetPlus{$\texttt{DMeet}^{+}$}

\newcommand{\acal}{\mathcal{A}}

\newcommand{\ecal}{\mathcal{E}}
\newcommand{\fcal}{\mathcal{F}}

\newcommand{\kcal}{\mathcal{K}}

\newcommand{\ocal}{\mathcal{O}}
\newcommand{\pcal}{\mathcal{P}}

\newcommand{\rcal}{\mathcal{R}}
\newcommand{\scal}{\mathcal{S}}

\newcommand{\defsymbol}{\stackrel{{\scriptscriptstyle \textup{\texttt{def}}}}{=}} 		

\global\long\def\eqdefII{\stackrel{{\scriptscriptstyle \mathrm{def}}}{=}}%
\global\long\def\eqdef{\eqdefII}%
\newcommand{\invack}[1]{\alpha_{{#1}}}

\newcommand{\Zinterval}[2]{ [#1.. #2]}

\newcommand{\inlineeq}[1]{\refstepcounter{equation}(\theequation)\ \(#1\)}

\newcounter{kproperty}
\newcommand{\kfrk}{\mathfrak{K}}
\newcommand{\theproperty}{\arabic{kproperty}}
\newcommand{\kproperty}[1]{\refstepcounter{kproperty}($\kfrk\theproperty$)\ \(#1\)}
\newcommand{\refkp}[1]{\ref{#1}}

\newcommand{\arrindex}[1]{%
  \ifstrempty{#1}{}{[#1]}%
}

\newenvironment{compactFigure}[1][htbp]{
  \begingroup
  \setlength\intextsep{8pt}
  \begin{figure}[#1]
}{
  \end{figure}
  \endgroup
}

\def\DMeetPlus{$\texttt{DMeet}^{+}$}
\def\GMeet{\texttt{GMeet}}
\def\GMeetPlus{$\texttt{GMeet}^+$}
\def\MonoBelow{\texttt{MonoBelow}}
\def\GMeetMono{\texttt{GMeetMono}}
\def\GMeetMonoLazy{\texttt{GMeetMonoLazy}}
\def\GMeetStar{$\GMeet{}^\star$}
\def\GMeetMonoStar{$\GMeetMono{}^\star$}
\def\sqleq{\sqsubseteq}
\def\sqgeq{\sqsupseteq}
\def\sqlt{\sqsubset}
\def\sqgt{\sqsupset}
\def\calF{\Fvar}
\def\joinF{\sqcup_{\scriptscriptstyle \calF}}
\def\meetF{\sqcap_{\scriptscriptstyle \calF}}

\def\meetE{\meetp{\jhspace{\Lat}}}

\def\calS{\mathcal{S}}
\def\joinS{\sqcup_{\scriptscriptstyle \calS}}

%% file: figures/knowledge-cmp1.pgf
\begingroup%
\makeatletter%
\begin{pgfpicture}%
\pgfpathrectangle{\pgfpointorigin}{\pgfqpoint{3.300000in}{2.500000in}}%
\pgfusepath{use as bounding box, clip}%
\begin{pgfscope}%
\pgfsetbuttcap%
\pgfsetmiterjoin%
\definecolor{currentfill}{rgb}{1.000000,1.000000,1.000000}%
\pgfsetfillcolor{currentfill}%
\pgfsetlinewidth{0.000000pt}%
\definecolor{currentstroke}{rgb}{1.000000,1.000000,1.000000}%
\pgfsetstrokecolor{currentstroke}%
\pgfsetdash{}{0pt}%
\pgfpathmoveto{\pgfqpoint{0.000000in}{0.000000in}}%
\pgfpathlineto{\pgfqpoint{3.300000in}{0.000000in}}%
\pgfpathlineto{\pgfqpoint{3.300000in}{2.500000in}}%
\pgfpathlineto{\pgfqpoint{0.000000in}{2.500000in}}%
\pgfpathclose%
\pgfusepath{fill}%
\end{pgfscope}%
\begin{pgfscope}%
\pgfsetbuttcap%
\pgfsetmiterjoin%
\definecolor{currentfill}{rgb}{1.000000,1.000000,1.000000}%
\pgfsetfillcolor{currentfill}%
\pgfsetlinewidth{0.000000pt}%
\definecolor{currentstroke}{rgb}{0.000000,0.000000,0.000000}%
\pgfsetstrokecolor{currentstroke}%
\pgfsetstrokeopacity{0.000000}%
\pgfsetdash{}{0pt}%
\pgfpathmoveto{\pgfqpoint{0.524236in}{0.502778in}}%
\pgfpathlineto{\pgfqpoint{3.089121in}{0.502778in}}%
\pgfpathlineto{\pgfqpoint{3.089121in}{2.378558in}}%
\pgfpathlineto{\pgfqpoint{0.524236in}{2.378558in}}%
\pgfpathclose%
\pgfusepath{fill}%
\end{pgfscope}%
\begin{pgfscope}%
\pgfpathrectangle{\pgfqpoint{0.524236in}{0.502778in}}{\pgfqpoint{2.564885in}{1.875780in}}%
\pgfusepath{clip}%
\pgfsetbuttcap%
\pgfsetroundjoin%
\pgfsetlinewidth{0.501875pt}%
\definecolor{currentstroke}{rgb}{0.000000,0.000000,0.000000}%
\pgfsetstrokecolor{currentstroke}%
\pgfsetstrokeopacity{0.500000}%
\pgfsetdash{{1.850000pt}{0.800000pt}}{0.000000pt}%
\pgfpathmoveto{\pgfqpoint{0.539443in}{0.502778in}}%
\pgfpathlineto{\pgfqpoint{0.539443in}{2.378558in}}%
\pgfusepath{stroke}%
\end{pgfscope}%
\begin{pgfscope}%
\pgfsetbuttcap%
\pgfsetroundjoin%
\definecolor{currentfill}{rgb}{0.000000,0.000000,0.000000}%
\pgfsetfillcolor{currentfill}%
\pgfsetlinewidth{0.803000pt}%
\definecolor{currentstroke}{rgb}{0.000000,0.000000,0.000000}%
\pgfsetstrokecolor{currentstroke}%
\pgfsetdash{}{0pt}%
\pgfsys@defobject{currentmarker}{\pgfqpoint{0.000000in}{-0.048611in}}{\pgfqpoint{0.000000in}{0.000000in}}{%
\pgfpathmoveto{\pgfqpoint{0.000000in}{0.000000in}}%
\pgfpathlineto{\pgfqpoint{0.000000in}{-0.048611in}}%
\pgfusepath{stroke,fill}%
}%
\begin{pgfscope}%
\pgfsys@transformshift{0.539443in}{0.502778in}%
\pgfsys@useobject{currentmarker}{}%
\end{pgfscope}%
\end{pgfscope}%
\begin{pgfscope}%
\definecolor{textcolor}{rgb}{0.000000,0.000000,0.000000}%
\pgfsetstrokecolor{textcolor}%
\pgfsetfillcolor{textcolor}%
\pgftext[x=0.539443in,y=0.405556in,,top]{\color{textcolor}\rmfamily\fontsize{8.000000}{9.600000}\selectfont 0}%
\end{pgfscope}%
\begin{pgfscope}%
\pgfpathrectangle{\pgfqpoint{0.524236in}{0.502778in}}{\pgfqpoint{2.564885in}{1.875780in}}%
\pgfusepath{clip}%
\pgfsetbuttcap%
\pgfsetroundjoin%
\pgfsetlinewidth{0.501875pt}%
\definecolor{currentstroke}{rgb}{0.000000,0.000000,0.000000}%
\pgfsetstrokecolor{currentstroke}%
\pgfsetstrokeopacity{0.500000}%
\pgfsetdash{{1.850000pt}{0.800000pt}}{0.000000pt}%
\pgfpathmoveto{\pgfqpoint{1.173061in}{0.502778in}}%
\pgfpathlineto{\pgfqpoint{1.173061in}{2.378558in}}%
\pgfusepath{stroke}%
\end{pgfscope}%
\begin{pgfscope}%
\pgfsetbuttcap%
\pgfsetroundjoin%
\definecolor{currentfill}{rgb}{0.000000,0.000000,0.000000}%
\pgfsetfillcolor{currentfill}%
\pgfsetlinewidth{0.803000pt}%
\definecolor{currentstroke}{rgb}{0.000000,0.000000,0.000000}%
\pgfsetstrokecolor{currentstroke}%
\pgfsetdash{}{0pt}%
\pgfsys@defobject{currentmarker}{\pgfqpoint{0.000000in}{-0.048611in}}{\pgfqpoint{0.000000in}{0.000000in}}{%
\pgfpathmoveto{\pgfqpoint{0.000000in}{0.000000in}}%
\pgfpathlineto{\pgfqpoint{0.000000in}{-0.048611in}}%
\pgfusepath{stroke,fill}%
}%
\begin{pgfscope}%
\pgfsys@transformshift{1.173061in}{0.502778in}%
\pgfsys@useobject{currentmarker}{}%
\end{pgfscope}%
\end{pgfscope}%
\begin{pgfscope}%
\definecolor{textcolor}{rgb}{0.000000,0.000000,0.000000}%
\pgfsetstrokecolor{textcolor}%
\pgfsetfillcolor{textcolor}%
\pgftext[x=1.173061in,y=0.405556in,,top]{\color{textcolor}\rmfamily\fontsize{8.000000}{9.600000}\selectfont 100}%
\end{pgfscope}%
\begin{pgfscope}%
\pgfpathrectangle{\pgfqpoint{0.524236in}{0.502778in}}{\pgfqpoint{2.564885in}{1.875780in}}%
\pgfusepath{clip}%
\pgfsetbuttcap%
\pgfsetroundjoin%
\pgfsetlinewidth{0.501875pt}%
\definecolor{currentstroke}{rgb}{0.000000,0.000000,0.000000}%
\pgfsetstrokecolor{currentstroke}%
\pgfsetstrokeopacity{0.500000}%
\pgfsetdash{{1.850000pt}{0.800000pt}}{0.000000pt}%
\pgfpathmoveto{\pgfqpoint{1.806679in}{0.502778in}}%
\pgfpathlineto{\pgfqpoint{1.806679in}{2.378558in}}%
\pgfusepath{stroke}%
\end{pgfscope}%
\begin{pgfscope}%
\pgfsetbuttcap%
\pgfsetroundjoin%
\definecolor{currentfill}{rgb}{0.000000,0.000000,0.000000}%
\pgfsetfillcolor{currentfill}%
\pgfsetlinewidth{0.803000pt}%
\definecolor{currentstroke}{rgb}{0.000000,0.000000,0.000000}%
\pgfsetstrokecolor{currentstroke}%
\pgfsetdash{}{0pt}%
\pgfsys@defobject{currentmarker}{\pgfqpoint{0.000000in}{-0.048611in}}{\pgfqpoint{0.000000in}{0.000000in}}{%
\pgfpathmoveto{\pgfqpoint{0.000000in}{0.000000in}}%
\pgfpathlineto{\pgfqpoint{0.000000in}{-0.048611in}}%
\pgfusepath{stroke,fill}%
}%
\begin{pgfscope}%
\pgfsys@transformshift{1.806679in}{0.502778in}%
\pgfsys@useobject{currentmarker}{}%
\end{pgfscope}%
\end{pgfscope}%
\begin{pgfscope}%
\definecolor{textcolor}{rgb}{0.000000,0.000000,0.000000}%
\pgfsetstrokecolor{textcolor}%
\pgfsetfillcolor{textcolor}%
\pgftext[x=1.806679in,y=0.405556in,,top]{\color{textcolor}\rmfamily\fontsize{8.000000}{9.600000}\selectfont 200}%
\end{pgfscope}%
\begin{pgfscope}%
\pgfpathrectangle{\pgfqpoint{0.524236in}{0.502778in}}{\pgfqpoint{2.564885in}{1.875780in}}%
\pgfusepath{clip}%
\pgfsetbuttcap%
\pgfsetroundjoin%
\pgfsetlinewidth{0.501875pt}%
\definecolor{currentstroke}{rgb}{0.000000,0.000000,0.000000}%
\pgfsetstrokecolor{currentstroke}%
\pgfsetstrokeopacity{0.500000}%
\pgfsetdash{{1.850000pt}{0.800000pt}}{0.000000pt}%
\pgfpathmoveto{\pgfqpoint{2.440297in}{0.502778in}}%
\pgfpathlineto{\pgfqpoint{2.440297in}{2.378558in}}%
\pgfusepath{stroke}%
\end{pgfscope}%
\begin{pgfscope}%
\pgfsetbuttcap%
\pgfsetroundjoin%
\definecolor{currentfill}{rgb}{0.000000,0.000000,0.000000}%
\pgfsetfillcolor{currentfill}%
\pgfsetlinewidth{0.803000pt}%
\definecolor{currentstroke}{rgb}{0.000000,0.000000,0.000000}%
\pgfsetstrokecolor{currentstroke}%
\pgfsetdash{}{0pt}%
\pgfsys@defobject{currentmarker}{\pgfqpoint{0.000000in}{-0.048611in}}{\pgfqpoint{0.000000in}{0.000000in}}{%
\pgfpathmoveto{\pgfqpoint{0.000000in}{0.000000in}}%
\pgfpathlineto{\pgfqpoint{0.000000in}{-0.048611in}}%
\pgfusepath{stroke,fill}%
}%
\begin{pgfscope}%
\pgfsys@transformshift{2.440297in}{0.502778in}%
\pgfsys@useobject{currentmarker}{}%
\end{pgfscope}%
\end{pgfscope}%
\begin{pgfscope}%
\definecolor{textcolor}{rgb}{0.000000,0.000000,0.000000}%
\pgfsetstrokecolor{textcolor}%
\pgfsetfillcolor{textcolor}%
\pgftext[x=2.440297in,y=0.405556in,,top]{\color{textcolor}\rmfamily\fontsize{8.000000}{9.600000}\selectfont 300}%
\end{pgfscope}%
\begin{pgfscope}%
\pgfpathrectangle{\pgfqpoint{0.524236in}{0.502778in}}{\pgfqpoint{2.564885in}{1.875780in}}%
\pgfusepath{clip}%
\pgfsetbuttcap%
\pgfsetroundjoin%
\pgfsetlinewidth{0.501875pt}%
\definecolor{currentstroke}{rgb}{0.000000,0.000000,0.000000}%
\pgfsetstrokecolor{currentstroke}%
\pgfsetstrokeopacity{0.500000}%
\pgfsetdash{{1.850000pt}{0.800000pt}}{0.000000pt}%
\pgfpathmoveto{\pgfqpoint{3.073915in}{0.502778in}}%
\pgfpathlineto{\pgfqpoint{3.073915in}{2.378558in}}%
\pgfusepath{stroke}%
\end{pgfscope}%
\begin{pgfscope}%
\pgfsetbuttcap%
\pgfsetroundjoin%
\definecolor{currentfill}{rgb}{0.000000,0.000000,0.000000}%
\pgfsetfillcolor{currentfill}%
\pgfsetlinewidth{0.803000pt}%
\definecolor{currentstroke}{rgb}{0.000000,0.000000,0.000000}%
\pgfsetstrokecolor{currentstroke}%
\pgfsetdash{}{0pt}%
\pgfsys@defobject{currentmarker}{\pgfqpoint{0.000000in}{-0.048611in}}{\pgfqpoint{0.000000in}{0.000000in}}{%
\pgfpathmoveto{\pgfqpoint{0.000000in}{0.000000in}}%
\pgfpathlineto{\pgfqpoint{0.000000in}{-0.048611in}}%
\pgfusepath{stroke,fill}%
}%
\begin{pgfscope}%
\pgfsys@transformshift{3.073915in}{0.502778in}%
\pgfsys@useobject{currentmarker}{}%
\end{pgfscope}%
\end{pgfscope}%
\begin{pgfscope}%
\definecolor{textcolor}{rgb}{0.000000,0.000000,0.000000}%
\pgfsetstrokecolor{textcolor}%
\pgfsetfillcolor{textcolor}%
\pgftext[x=3.073915in,y=0.405556in,,top]{\color{textcolor}\rmfamily\fontsize{8.000000}{9.600000}\selectfont 400}%
\end{pgfscope}%
\begin{pgfscope}%
\definecolor{textcolor}{rgb}{0.000000,0.000000,0.000000}%
\pgfsetstrokecolor{textcolor}%
\pgfsetfillcolor{textcolor}%
\pgftext[x=1.806679in,y=0.251234in,,top]{\color{textcolor}\rmfamily\fontsize{8.000000}{9.600000}\selectfont \(\displaystyle |\Omega|\)}%
\end{pgfscope}%
\begin{pgfscope}%
\pgfpathrectangle{\pgfqpoint{0.524236in}{0.502778in}}{\pgfqpoint{2.564885in}{1.875780in}}%
\pgfusepath{clip}%
\pgfsetbuttcap%
\pgfsetroundjoin%
\pgfsetlinewidth{0.501875pt}%
\definecolor{currentstroke}{rgb}{0.000000,0.000000,0.000000}%
\pgfsetstrokecolor{currentstroke}%
\pgfsetstrokeopacity{0.500000}%
\pgfsetdash{{1.850000pt}{0.800000pt}}{0.000000pt}%
\pgfpathmoveto{\pgfqpoint{0.524236in}{0.587770in}}%
\pgfpathlineto{\pgfqpoint{3.089121in}{0.587770in}}%
\pgfusepath{stroke}%
\end{pgfscope}%
\begin{pgfscope}%
\pgfsetbuttcap%
\pgfsetroundjoin%
\definecolor{currentfill}{rgb}{0.000000,0.000000,0.000000}%
\pgfsetfillcolor{currentfill}%
\pgfsetlinewidth{0.803000pt}%
\definecolor{currentstroke}{rgb}{0.000000,0.000000,0.000000}%
\pgfsetstrokecolor{currentstroke}%
\pgfsetdash{}{0pt}%
\pgfsys@defobject{currentmarker}{\pgfqpoint{-0.048611in}{0.000000in}}{\pgfqpoint{0.000000in}{0.000000in}}{%
\pgfpathmoveto{\pgfqpoint{0.000000in}{0.000000in}}%
\pgfpathlineto{\pgfqpoint{-0.048611in}{0.000000in}}%
\pgfusepath{stroke,fill}%
}%
\begin{pgfscope}%
\pgfsys@transformshift{0.524236in}{0.587770in}%
\pgfsys@useobject{currentmarker}{}%
\end{pgfscope}%
\end{pgfscope}%
\begin{pgfscope}%
\definecolor{textcolor}{rgb}{0.000000,0.000000,0.000000}%
\pgfsetstrokecolor{textcolor}%
\pgfsetfillcolor{textcolor}%
\pgftext[x=0.367985in, y=0.549189in, left, base]{\color{textcolor}\rmfamily\fontsize{8.000000}{9.600000}\selectfont 0}%
\end{pgfscope}%
\begin{pgfscope}%
\pgfpathrectangle{\pgfqpoint{0.524236in}{0.502778in}}{\pgfqpoint{2.564885in}{1.875780in}}%
\pgfusepath{clip}%
\pgfsetbuttcap%
\pgfsetroundjoin%
\pgfsetlinewidth{0.501875pt}%
\definecolor{currentstroke}{rgb}{0.000000,0.000000,0.000000}%
\pgfsetstrokecolor{currentstroke}%
\pgfsetstrokeopacity{0.500000}%
\pgfsetdash{{1.850000pt}{0.800000pt}}{0.000000pt}%
\pgfpathmoveto{\pgfqpoint{0.524236in}{0.937601in}}%
\pgfpathlineto{\pgfqpoint{3.089121in}{0.937601in}}%
\pgfusepath{stroke}%
\end{pgfscope}%
\begin{pgfscope}%
\pgfsetbuttcap%
\pgfsetroundjoin%
\definecolor{currentfill}{rgb}{0.000000,0.000000,0.000000}%
\pgfsetfillcolor{currentfill}%
\pgfsetlinewidth{0.803000pt}%
\definecolor{currentstroke}{rgb}{0.000000,0.000000,0.000000}%
\pgfsetstrokecolor{currentstroke}%
\pgfsetdash{}{0pt}%
\pgfsys@defobject{currentmarker}{\pgfqpoint{-0.048611in}{0.000000in}}{\pgfqpoint{0.000000in}{0.000000in}}{%
\pgfpathmoveto{\pgfqpoint{0.000000in}{0.000000in}}%
\pgfpathlineto{\pgfqpoint{-0.048611in}{0.000000in}}%
\pgfusepath{stroke,fill}%
}%
\begin{pgfscope}%
\pgfsys@transformshift{0.524236in}{0.937601in}%
\pgfsys@useobject{currentmarker}{}%
\end{pgfscope}%
\end{pgfscope}%
\begin{pgfscope}%
\definecolor{textcolor}{rgb}{0.000000,0.000000,0.000000}%
\pgfsetstrokecolor{textcolor}%
\pgfsetfillcolor{textcolor}%
\pgftext[x=0.308957in, y=0.899021in, left, base]{\color{textcolor}\rmfamily\fontsize{8.000000}{9.600000}\selectfont 10}%
\end{pgfscope}%
\begin{pgfscope}%
\pgfpathrectangle{\pgfqpoint{0.524236in}{0.502778in}}{\pgfqpoint{2.564885in}{1.875780in}}%
\pgfusepath{clip}%
\pgfsetbuttcap%
\pgfsetroundjoin%
\pgfsetlinewidth{0.501875pt}%
\definecolor{currentstroke}{rgb}{0.000000,0.000000,0.000000}%
\pgfsetstrokecolor{currentstroke}%
\pgfsetstrokeopacity{0.500000}%
\pgfsetdash{{1.850000pt}{0.800000pt}}{0.000000pt}%
\pgfpathmoveto{\pgfqpoint{0.524236in}{1.287432in}}%
\pgfpathlineto{\pgfqpoint{3.089121in}{1.287432in}}%
\pgfusepath{stroke}%
\end{pgfscope}%
\begin{pgfscope}%
\pgfsetbuttcap%
\pgfsetroundjoin%
\definecolor{currentfill}{rgb}{0.000000,0.000000,0.000000}%
\pgfsetfillcolor{currentfill}%
\pgfsetlinewidth{0.803000pt}%
\definecolor{currentstroke}{rgb}{0.000000,0.000000,0.000000}%
\pgfsetstrokecolor{currentstroke}%
\pgfsetdash{}{0pt}%
\pgfsys@defobject{currentmarker}{\pgfqpoint{-0.048611in}{0.000000in}}{\pgfqpoint{0.000000in}{0.000000in}}{%
\pgfpathmoveto{\pgfqpoint{0.000000in}{0.000000in}}%
\pgfpathlineto{\pgfqpoint{-0.048611in}{0.000000in}}%
\pgfusepath{stroke,fill}%
}%
\begin{pgfscope}%
\pgfsys@transformshift{0.524236in}{1.287432in}%
\pgfsys@useobject{currentmarker}{}%
\end{pgfscope}%
\end{pgfscope}%
\begin{pgfscope}%
\definecolor{textcolor}{rgb}{0.000000,0.000000,0.000000}%
\pgfsetstrokecolor{textcolor}%
\pgfsetfillcolor{textcolor}%
\pgftext[x=0.308957in, y=1.248852in, left, base]{\color{textcolor}\rmfamily\fontsize{8.000000}{9.600000}\selectfont 20}%
\end{pgfscope}%
\begin{pgfscope}%
\pgfpathrectangle{\pgfqpoint{0.524236in}{0.502778in}}{\pgfqpoint{2.564885in}{1.875780in}}%
\pgfusepath{clip}%
\pgfsetbuttcap%
\pgfsetroundjoin%
\pgfsetlinewidth{0.501875pt}%
\definecolor{currentstroke}{rgb}{0.000000,0.000000,0.000000}%
\pgfsetstrokecolor{currentstroke}%
\pgfsetstrokeopacity{0.500000}%
\pgfsetdash{{1.850000pt}{0.800000pt}}{0.000000pt}%
\pgfpathmoveto{\pgfqpoint{0.524236in}{1.637264in}}%
\pgfpathlineto{\pgfqpoint{3.089121in}{1.637264in}}%
\pgfusepath{stroke}%
\end{pgfscope}%
\begin{pgfscope}%
\pgfsetbuttcap%
\pgfsetroundjoin%
\definecolor{currentfill}{rgb}{0.000000,0.000000,0.000000}%
\pgfsetfillcolor{currentfill}%
\pgfsetlinewidth{0.803000pt}%
\definecolor{currentstroke}{rgb}{0.000000,0.000000,0.000000}%
\pgfsetstrokecolor{currentstroke}%
\pgfsetdash{}{0pt}%
\pgfsys@defobject{currentmarker}{\pgfqpoint{-0.048611in}{0.000000in}}{\pgfqpoint{0.000000in}{0.000000in}}{%
\pgfpathmoveto{\pgfqpoint{0.000000in}{0.000000in}}%
\pgfpathlineto{\pgfqpoint{-0.048611in}{0.000000in}}%
\pgfusepath{stroke,fill}%
}%
\begin{pgfscope}%
\pgfsys@transformshift{0.524236in}{1.637264in}%
\pgfsys@useobject{currentmarker}{}%
\end{pgfscope}%
\end{pgfscope}%
\begin{pgfscope}%
\definecolor{textcolor}{rgb}{0.000000,0.000000,0.000000}%
\pgfsetstrokecolor{textcolor}%
\pgfsetfillcolor{textcolor}%
\pgftext[x=0.308957in, y=1.598684in, left, base]{\color{textcolor}\rmfamily\fontsize{8.000000}{9.600000}\selectfont 30}%
\end{pgfscope}%
\begin{pgfscope}%
\pgfpathrectangle{\pgfqpoint{0.524236in}{0.502778in}}{\pgfqpoint{2.564885in}{1.875780in}}%
\pgfusepath{clip}%
\pgfsetbuttcap%
\pgfsetroundjoin%
\pgfsetlinewidth{0.501875pt}%
\definecolor{currentstroke}{rgb}{0.000000,0.000000,0.000000}%
\pgfsetstrokecolor{currentstroke}%
\pgfsetstrokeopacity{0.500000}%
\pgfsetdash{{1.850000pt}{0.800000pt}}{0.000000pt}%
\pgfpathmoveto{\pgfqpoint{0.524236in}{1.987095in}}%
\pgfpathlineto{\pgfqpoint{3.089121in}{1.987095in}}%
\pgfusepath{stroke}%
\end{pgfscope}%
\begin{pgfscope}%
\pgfsetbuttcap%
\pgfsetroundjoin%
\definecolor{currentfill}{rgb}{0.000000,0.000000,0.000000}%
\pgfsetfillcolor{currentfill}%
\pgfsetlinewidth{0.803000pt}%
\definecolor{currentstroke}{rgb}{0.000000,0.000000,0.000000}%
\pgfsetstrokecolor{currentstroke}%
\pgfsetdash{}{0pt}%
\pgfsys@defobject{currentmarker}{\pgfqpoint{-0.048611in}{0.000000in}}{\pgfqpoint{0.000000in}{0.000000in}}{%
\pgfpathmoveto{\pgfqpoint{0.000000in}{0.000000in}}%
\pgfpathlineto{\pgfqpoint{-0.048611in}{0.000000in}}%
\pgfusepath{stroke,fill}%
}%
\begin{pgfscope}%
\pgfsys@transformshift{0.524236in}{1.987095in}%
\pgfsys@useobject{currentmarker}{}%
\end{pgfscope}%
\end{pgfscope}%
\begin{pgfscope}%
\definecolor{textcolor}{rgb}{0.000000,0.000000,0.000000}%
\pgfsetstrokecolor{textcolor}%
\pgfsetfillcolor{textcolor}%
\pgftext[x=0.308957in, y=1.948515in, left, base]{\color{textcolor}\rmfamily\fontsize{8.000000}{9.600000}\selectfont 40}%
\end{pgfscope}%
\begin{pgfscope}%
\pgfpathrectangle{\pgfqpoint{0.524236in}{0.502778in}}{\pgfqpoint{2.564885in}{1.875780in}}%
\pgfusepath{clip}%
\pgfsetbuttcap%
\pgfsetroundjoin%
\pgfsetlinewidth{0.501875pt}%
\definecolor{currentstroke}{rgb}{0.000000,0.000000,0.000000}%
\pgfsetstrokecolor{currentstroke}%
\pgfsetstrokeopacity{0.500000}%
\pgfsetdash{{1.850000pt}{0.800000pt}}{0.000000pt}%
\pgfpathmoveto{\pgfqpoint{0.524236in}{2.336927in}}%
\pgfpathlineto{\pgfqpoint{3.089121in}{2.336927in}}%
\pgfusepath{stroke}%
\end{pgfscope}%
\begin{pgfscope}%
\pgfsetbuttcap%
\pgfsetroundjoin%
\definecolor{currentfill}{rgb}{0.000000,0.000000,0.000000}%
\pgfsetfillcolor{currentfill}%
\pgfsetlinewidth{0.803000pt}%
\definecolor{currentstroke}{rgb}{0.000000,0.000000,0.000000}%
\pgfsetstrokecolor{currentstroke}%
\pgfsetdash{}{0pt}%
\pgfsys@defobject{currentmarker}{\pgfqpoint{-0.048611in}{0.000000in}}{\pgfqpoint{0.000000in}{0.000000in}}{%
\pgfpathmoveto{\pgfqpoint{0.000000in}{0.000000in}}%
\pgfpathlineto{\pgfqpoint{-0.048611in}{0.000000in}}%
\pgfusepath{stroke,fill}%
}%
\begin{pgfscope}%
\pgfsys@transformshift{0.524236in}{2.336927in}%
\pgfsys@useobject{currentmarker}{}%
\end{pgfscope}%
\end{pgfscope}%
\begin{pgfscope}%
\definecolor{textcolor}{rgb}{0.000000,0.000000,0.000000}%
\pgfsetstrokecolor{textcolor}%
\pgfsetfillcolor{textcolor}%
\pgftext[x=0.308957in, y=2.298346in, left, base]{\color{textcolor}\rmfamily\fontsize{8.000000}{9.600000}\selectfont 50}%
\end{pgfscope}%
\begin{pgfscope}%
\definecolor{textcolor}{rgb}{0.000000,0.000000,0.000000}%
\pgfsetstrokecolor{textcolor}%
\pgfsetfillcolor{textcolor}%
\pgftext[x=0.253401in,y=1.440668in,,bottom,rotate=90.000000]{\color{textcolor}\rmfamily\fontsize{8.000000}{9.600000}\selectfont Average runtime [ms]}%
\end{pgfscope}%
\begin{pgfscope}%
\pgfpathrectangle{\pgfqpoint{0.524236in}{0.502778in}}{\pgfqpoint{2.564885in}{1.875780in}}%
\pgfusepath{clip}%
\pgfsetrectcap%
\pgfsetroundjoin%
\pgfsetlinewidth{1.505625pt}%
\definecolor{currentstroke}{rgb}{0.000000,0.000000,0.000000}%
\pgfsetstrokecolor{currentstroke}%
\pgfsetdash{}{0pt}%
\pgfpathmoveto{\pgfqpoint{0.640822in}{0.589147in}}%
\pgfpathlineto{\pgfqpoint{0.691511in}{0.589701in}}%
\pgfpathlineto{\pgfqpoint{0.742201in}{0.590221in}}%
\pgfpathlineto{\pgfqpoint{0.843580in}{0.591508in}}%
\pgfpathlineto{\pgfqpoint{0.944958in}{0.591806in}}%
\pgfpathlineto{\pgfqpoint{1.147716in}{0.593712in}}%
\pgfpathlineto{\pgfqpoint{1.350474in}{0.595992in}}%
\pgfpathlineto{\pgfqpoint{1.755989in}{0.599761in}}%
\pgfpathlineto{\pgfqpoint{2.161505in}{0.599730in}}%
\pgfpathlineto{\pgfqpoint{2.972536in}{0.606414in}}%
\pgfusepath{stroke}%
\end{pgfscope}%
\begin{pgfscope}%
\pgfpathrectangle{\pgfqpoint{0.524236in}{0.502778in}}{\pgfqpoint{2.564885in}{1.875780in}}%
\pgfusepath{clip}%
\pgfsetbuttcap%
\pgfsetroundjoin%
\pgfsetlinewidth{1.505625pt}%
\definecolor{currentstroke}{rgb}{1.000000,0.000000,0.000000}%
\pgfsetstrokecolor{currentstroke}%
\pgfsetdash{{5.550000pt}{2.400000pt}}{0.000000pt}%
\pgfpathmoveto{\pgfqpoint{0.640822in}{0.588041in}}%
\pgfpathlineto{\pgfqpoint{0.691511in}{0.588132in}}%
\pgfpathlineto{\pgfqpoint{0.742201in}{0.588248in}}%
\pgfpathlineto{\pgfqpoint{0.843580in}{0.588481in}}%
\pgfpathlineto{\pgfqpoint{0.944958in}{0.588514in}}%
\pgfpathlineto{\pgfqpoint{1.147716in}{0.588885in}}%
\pgfpathlineto{\pgfqpoint{1.350474in}{0.589272in}}%
\pgfpathlineto{\pgfqpoint{1.755989in}{0.589781in}}%
\pgfpathlineto{\pgfqpoint{2.161505in}{0.589759in}}%
\pgfpathlineto{\pgfqpoint{2.972536in}{0.591543in}}%
\pgfusepath{stroke}%
\end{pgfscope}%
\begin{pgfscope}%
\pgfpathrectangle{\pgfqpoint{0.524236in}{0.502778in}}{\pgfqpoint{2.564885in}{1.875780in}}%
\pgfusepath{clip}%
\pgfsetbuttcap%
\pgfsetroundjoin%
\pgfsetlinewidth{1.505625pt}%
\definecolor{currentstroke}{rgb}{0.000000,0.000000,1.000000}%
\pgfsetstrokecolor{currentstroke}%
\pgfsetdash{{1.500000pt}{2.475000pt}}{0.000000pt}%
\pgfpathmoveto{\pgfqpoint{0.640822in}{0.589398in}}%
\pgfpathlineto{\pgfqpoint{0.691511in}{0.591044in}}%
\pgfpathlineto{\pgfqpoint{0.742201in}{0.593927in}}%
\pgfpathlineto{\pgfqpoint{0.843580in}{0.601825in}}%
\pgfpathlineto{\pgfqpoint{0.944958in}{0.610904in}}%
\pgfpathlineto{\pgfqpoint{1.147716in}{0.636626in}}%
\pgfpathlineto{\pgfqpoint{1.350474in}{0.685391in}}%
\pgfpathlineto{\pgfqpoint{1.755989in}{0.778233in}}%
\pgfpathlineto{\pgfqpoint{2.161505in}{0.847766in}}%
\pgfpathlineto{\pgfqpoint{2.972536in}{1.225212in}}%
\pgfusepath{stroke}%
\end{pgfscope}%
\begin{pgfscope}%
\pgfpathrectangle{\pgfqpoint{0.524236in}{0.502778in}}{\pgfqpoint{2.564885in}{1.875780in}}%
\pgfusepath{clip}%
\pgfsetbuttcap%
\pgfsetroundjoin%
\pgfsetlinewidth{1.505625pt}%
\definecolor{currentstroke}{rgb}{0.000000,0.500000,0.000000}%
\pgfsetstrokecolor{currentstroke}%
\pgfsetdash{{9.600000pt}{2.400000pt}{1.500000pt}{2.400000pt}}{0.000000pt}%
\pgfpathmoveto{\pgfqpoint{0.640822in}{0.592786in}}%
\pgfpathlineto{\pgfqpoint{0.691511in}{0.597628in}}%
\pgfpathlineto{\pgfqpoint{0.742201in}{0.604693in}}%
\pgfpathlineto{\pgfqpoint{0.843580in}{0.626403in}}%
\pgfpathlineto{\pgfqpoint{0.944958in}{0.646157in}}%
\pgfpathlineto{\pgfqpoint{1.147716in}{0.716511in}}%
\pgfpathlineto{\pgfqpoint{1.350474in}{0.842595in}}%
\pgfpathlineto{\pgfqpoint{1.755989in}{1.152394in}}%
\pgfpathlineto{\pgfqpoint{2.161505in}{1.306941in}}%
\pgfpathlineto{\pgfqpoint{2.972536in}{2.293295in}}%
\pgfusepath{stroke}%
\end{pgfscope}%
\begin{pgfscope}%
\pgfsetrectcap%
\pgfsetmiterjoin%
\pgfsetlinewidth{0.803000pt}%
\definecolor{currentstroke}{rgb}{0.000000,0.000000,0.000000}%
\pgfsetstrokecolor{currentstroke}%
\pgfsetdash{}{0pt}%
\pgfpathmoveto{\pgfqpoint{0.524236in}{0.502778in}}%
\pgfpathlineto{\pgfqpoint{0.524236in}{2.378558in}}%
\pgfusepath{stroke}%
\end{pgfscope}%
\begin{pgfscope}%
\pgfsetrectcap%
\pgfsetmiterjoin%
\pgfsetlinewidth{0.803000pt}%
\definecolor{currentstroke}{rgb}{0.000000,0.000000,0.000000}%
\pgfsetstrokecolor{currentstroke}%
\pgfsetdash{}{0pt}%
\pgfpathmoveto{\pgfqpoint{3.089121in}{0.502778in}}%
\pgfpathlineto{\pgfqpoint{3.089121in}{2.378558in}}%
\pgfusepath{stroke}%
\end{pgfscope}%
\begin{pgfscope}%
\pgfsetrectcap%
\pgfsetmiterjoin%
\pgfsetlinewidth{0.803000pt}%
\definecolor{currentstroke}{rgb}{0.000000,0.000000,0.000000}%
\pgfsetstrokecolor{currentstroke}%
\pgfsetdash{}{0pt}%
\pgfpathmoveto{\pgfqpoint{0.524236in}{0.502778in}}%
\pgfpathlineto{\pgfqpoint{3.089121in}{0.502778in}}%
\pgfusepath{stroke}%
\end{pgfscope}%
\begin{pgfscope}%
\pgfsetrectcap%
\pgfsetmiterjoin%
\pgfsetlinewidth{0.803000pt}%
\definecolor{currentstroke}{rgb}{0.000000,0.000000,0.000000}%
\pgfsetstrokecolor{currentstroke}%
\pgfsetdash{}{0pt}%
\pgfpathmoveto{\pgfqpoint{0.524236in}{2.378558in}}%
\pgfpathlineto{\pgfqpoint{3.089121in}{2.378558in}}%
\pgfusepath{stroke}%
\end{pgfscope}%
\begin{pgfscope}%
\pgfsetbuttcap%
\pgfsetmiterjoin%
\definecolor{currentfill}{rgb}{1.000000,1.000000,1.000000}%
\pgfsetfillcolor{currentfill}%
\pgfsetlinewidth{1.003750pt}%
\definecolor{currentstroke}{rgb}{0.800000,0.800000,0.800000}%
\pgfsetstrokecolor{currentstroke}%
\pgfsetdash{}{0pt}%
\pgfpathmoveto{\pgfqpoint{0.602014in}{1.669916in}}%
\pgfpathlineto{\pgfqpoint{2.030000in}{1.669916in}}%
\pgfpathquadraticcurveto{\pgfqpoint{2.052222in}{1.669916in}}{\pgfqpoint{2.052222in}{1.692138in}}%
\pgfpathlineto{\pgfqpoint{2.052222in}{2.300780in}}%
\pgfpathquadraticcurveto{\pgfqpoint{2.052222in}{2.323003in}}{\pgfqpoint{2.030000in}{2.323003in}}%
\pgfpathlineto{\pgfqpoint{0.602014in}{2.323003in}}%
\pgfpathquadraticcurveto{\pgfqpoint{0.579792in}{2.323003in}}{\pgfqpoint{0.579792in}{2.300780in}}%
\pgfpathlineto{\pgfqpoint{0.579792in}{1.692138in}}%
\pgfpathquadraticcurveto{\pgfqpoint{0.579792in}{1.669916in}}{\pgfqpoint{0.602014in}{1.669916in}}%
\pgfpathclose%
\pgfusepath{stroke,fill}%
\end{pgfscope}%
\begin{pgfscope}%
\pgfsetrectcap%
\pgfsetroundjoin%
\pgfsetlinewidth{1.505625pt}%
\definecolor{currentstroke}{rgb}{0.000000,0.000000,0.000000}%
\pgfsetstrokecolor{currentstroke}%
\pgfsetdash{}{0pt}%
\pgfpathmoveto{\pgfqpoint{0.624236in}{2.239669in}}%
\pgfpathlineto{\pgfqpoint{0.846458in}{2.239669in}}%
\pgfusepath{stroke}%
\end{pgfscope}%
\begin{pgfscope}%
\definecolor{textcolor}{rgb}{0.000000,0.000000,0.000000}%
\pgfsetstrokecolor{textcolor}%
\pgfsetfillcolor{textcolor}%
\pgftext[x=0.935347in,y=2.200780in,left,base]{\color{textcolor}\rmfamily\fontsize{8.000000}{9.600000}\selectfont DisjointSet}%
\end{pgfscope}%
\begin{pgfscope}%
\pgfsetbuttcap%
\pgfsetroundjoin%
\pgfsetlinewidth{1.505625pt}%
\definecolor{currentstroke}{rgb}{1.000000,0.000000,0.000000}%
\pgfsetstrokecolor{currentstroke}%
\pgfsetdash{{5.550000pt}{2.400000pt}}{0.000000pt}%
\pgfpathmoveto{\pgfqpoint{0.624236in}{2.084731in}}%
\pgfpathlineto{\pgfqpoint{0.846458in}{2.084731in}}%
\pgfusepath{stroke}%
\end{pgfscope}%
\begin{pgfscope}%
\definecolor{textcolor}{rgb}{0.000000,0.000000,0.000000}%
\pgfsetstrokecolor{textcolor}%
\pgfsetfillcolor{textcolor}%
\pgftext[x=0.935347in,y=2.045842in,left,base]{\color{textcolor}\rmfamily\fontsize{8.000000}{9.600000}\selectfont Cached operator}%
\end{pgfscope}%
\begin{pgfscope}%
\pgfsetbuttcap%
\pgfsetroundjoin%
\pgfsetlinewidth{1.505625pt}%
\definecolor{currentstroke}{rgb}{0.000000,0.000000,1.000000}%
\pgfsetstrokecolor{currentstroke}%
\pgfsetdash{{1.500000pt}{2.475000pt}}{0.000000pt}%
\pgfpathmoveto{\pgfqpoint{0.624236in}{1.929793in}}%
\pgfpathlineto{\pgfqpoint{0.846458in}{1.929793in}}%
\pgfusepath{stroke}%
\end{pgfscope}%
\begin{pgfscope}%
\definecolor{textcolor}{rgb}{0.000000,0.000000,0.000000}%
\pgfsetstrokecolor{textcolor}%
\pgfsetfillcolor{textcolor}%
\pgftext[x=0.935347in,y=1.890904in,left,base]{\color{textcolor}\rmfamily\fontsize{8.000000}{9.600000}\selectfont Relation}%
\end{pgfscope}%
\begin{pgfscope}%
\pgfsetbuttcap%
\pgfsetroundjoin%
\pgfsetlinewidth{1.505625pt}%
\definecolor{currentstroke}{rgb}{0.000000,0.500000,0.000000}%
\pgfsetstrokecolor{currentstroke}%
\pgfsetdash{{9.600000pt}{2.400000pt}{1.500000pt}{2.400000pt}}{0.000000pt}%
\pgfpathmoveto{\pgfqpoint{0.624236in}{1.774854in}}%
\pgfpathlineto{\pgfqpoint{0.846458in}{1.774854in}}%
\pgfusepath{stroke}%
\end{pgfscope}%
\begin{pgfscope}%
\definecolor{textcolor}{rgb}{0.000000,0.000000,0.000000}%
\pgfsetstrokecolor{textcolor}%
\pgfsetfillcolor{textcolor}%
\pgftext[x=0.935347in,y=1.735965in,left,base]{\color{textcolor}\rmfamily\fontsize{8.000000}{9.600000}\selectfont Non-cached operator}%
\end{pgfscope}%
\end{pgfpicture}%
\makeatother%
\endgroup%

%% file: figures/knowledge-cmp2.pgf
\begingroup%
\makeatletter%
\begin{pgfpicture}%
\pgfpathrectangle{\pgfpointorigin}{\pgfqpoint{3.300000in}{2.500000in}}%
\pgfusepath{use as bounding box, clip}%
\begin{pgfscope}%
\pgfsetbuttcap%
\pgfsetmiterjoin%
\definecolor{currentfill}{rgb}{1.000000,1.000000,1.000000}%
\pgfsetfillcolor{currentfill}%
\pgfsetlinewidth{0.000000pt}%
\definecolor{currentstroke}{rgb}{1.000000,1.000000,1.000000}%
\pgfsetstrokecolor{currentstroke}%
\pgfsetdash{}{0pt}%
\pgfpathmoveto{\pgfqpoint{0.000000in}{0.000000in}}%
\pgfpathlineto{\pgfqpoint{3.300000in}{0.000000in}}%
\pgfpathlineto{\pgfqpoint{3.300000in}{2.500000in}}%
\pgfpathlineto{\pgfqpoint{0.000000in}{2.500000in}}%
\pgfpathclose%
\pgfusepath{fill}%
\end{pgfscope}%
\begin{pgfscope}%
\pgfsetbuttcap%
\pgfsetmiterjoin%
\definecolor{currentfill}{rgb}{1.000000,1.000000,1.000000}%
\pgfsetfillcolor{currentfill}%
\pgfsetlinewidth{0.000000pt}%
\definecolor{currentstroke}{rgb}{0.000000,0.000000,0.000000}%
\pgfsetstrokecolor{currentstroke}%
\pgfsetstrokeopacity{0.000000}%
\pgfsetdash{}{0pt}%
\pgfpathmoveto{\pgfqpoint{0.559653in}{0.502778in}}%
\pgfpathlineto{\pgfqpoint{3.089121in}{0.502778in}}%
\pgfpathlineto{\pgfqpoint{3.089121in}{2.380000in}}%
\pgfpathlineto{\pgfqpoint{0.559653in}{2.380000in}}%
\pgfpathclose%
\pgfusepath{fill}%
\end{pgfscope}%
\begin{pgfscope}%
\pgfpathrectangle{\pgfqpoint{0.559653in}{0.502778in}}{\pgfqpoint{2.529469in}{1.877222in}}%
\pgfusepath{clip}%
\pgfsetbuttcap%
\pgfsetroundjoin%
\pgfsetlinewidth{0.501875pt}%
\definecolor{currentstroke}{rgb}{0.000000,0.000000,0.000000}%
\pgfsetstrokecolor{currentstroke}%
\pgfsetstrokeopacity{0.500000}%
\pgfsetdash{{1.850000pt}{0.800000pt}}{0.000000pt}%
\pgfpathmoveto{\pgfqpoint{0.574650in}{0.502778in}}%
\pgfpathlineto{\pgfqpoint{0.574650in}{2.380000in}}%
\pgfusepath{stroke}%
\end{pgfscope}%
\begin{pgfscope}%
\pgfsetbuttcap%
\pgfsetroundjoin%
\definecolor{currentfill}{rgb}{0.000000,0.000000,0.000000}%
\pgfsetfillcolor{currentfill}%
\pgfsetlinewidth{0.803000pt}%
\definecolor{currentstroke}{rgb}{0.000000,0.000000,0.000000}%
\pgfsetstrokecolor{currentstroke}%
\pgfsetdash{}{0pt}%
\pgfsys@defobject{currentmarker}{\pgfqpoint{0.000000in}{-0.048611in}}{\pgfqpoint{0.000000in}{0.000000in}}{%
\pgfpathmoveto{\pgfqpoint{0.000000in}{0.000000in}}%
\pgfpathlineto{\pgfqpoint{0.000000in}{-0.048611in}}%
\pgfusepath{stroke,fill}%
}%
\begin{pgfscope}%
\pgfsys@transformshift{0.574650in}{0.502778in}%
\pgfsys@useobject{currentmarker}{}%
\end{pgfscope}%
\end{pgfscope}%
\begin{pgfscope}%
\definecolor{textcolor}{rgb}{0.000000,0.000000,0.000000}%
\pgfsetstrokecolor{textcolor}%
\pgfsetfillcolor{textcolor}%
\pgftext[x=0.574650in,y=0.405556in,,top]{\color{textcolor}\rmfamily\fontsize{8.000000}{9.600000}\selectfont 0}%
\end{pgfscope}%
\begin{pgfscope}%
\pgfpathrectangle{\pgfqpoint{0.559653in}{0.502778in}}{\pgfqpoint{2.529469in}{1.877222in}}%
\pgfusepath{clip}%
\pgfsetbuttcap%
\pgfsetroundjoin%
\pgfsetlinewidth{0.501875pt}%
\definecolor{currentstroke}{rgb}{0.000000,0.000000,0.000000}%
\pgfsetstrokecolor{currentstroke}%
\pgfsetstrokeopacity{0.500000}%
\pgfsetdash{{1.850000pt}{0.800000pt}}{0.000000pt}%
\pgfpathmoveto{\pgfqpoint{1.199518in}{0.502778in}}%
\pgfpathlineto{\pgfqpoint{1.199518in}{2.380000in}}%
\pgfusepath{stroke}%
\end{pgfscope}%
\begin{pgfscope}%
\pgfsetbuttcap%
\pgfsetroundjoin%
\definecolor{currentfill}{rgb}{0.000000,0.000000,0.000000}%
\pgfsetfillcolor{currentfill}%
\pgfsetlinewidth{0.803000pt}%
\definecolor{currentstroke}{rgb}{0.000000,0.000000,0.000000}%
\pgfsetstrokecolor{currentstroke}%
\pgfsetdash{}{0pt}%
\pgfsys@defobject{currentmarker}{\pgfqpoint{0.000000in}{-0.048611in}}{\pgfqpoint{0.000000in}{0.000000in}}{%
\pgfpathmoveto{\pgfqpoint{0.000000in}{0.000000in}}%
\pgfpathlineto{\pgfqpoint{0.000000in}{-0.048611in}}%
\pgfusepath{stroke,fill}%
}%
\begin{pgfscope}%
\pgfsys@transformshift{1.199518in}{0.502778in}%
\pgfsys@useobject{currentmarker}{}%
\end{pgfscope}%
\end{pgfscope}%
\begin{pgfscope}%
\definecolor{textcolor}{rgb}{0.000000,0.000000,0.000000}%
\pgfsetstrokecolor{textcolor}%
\pgfsetfillcolor{textcolor}%
\pgftext[x=1.199518in,y=0.405556in,,top]{\color{textcolor}\rmfamily\fontsize{8.000000}{9.600000}\selectfont 100}%
\end{pgfscope}%
\begin{pgfscope}%
\pgfpathrectangle{\pgfqpoint{0.559653in}{0.502778in}}{\pgfqpoint{2.529469in}{1.877222in}}%
\pgfusepath{clip}%
\pgfsetbuttcap%
\pgfsetroundjoin%
\pgfsetlinewidth{0.501875pt}%
\definecolor{currentstroke}{rgb}{0.000000,0.000000,0.000000}%
\pgfsetstrokecolor{currentstroke}%
\pgfsetstrokeopacity{0.500000}%
\pgfsetdash{{1.850000pt}{0.800000pt}}{0.000000pt}%
\pgfpathmoveto{\pgfqpoint{1.824387in}{0.502778in}}%
\pgfpathlineto{\pgfqpoint{1.824387in}{2.380000in}}%
\pgfusepath{stroke}%
\end{pgfscope}%
\begin{pgfscope}%
\pgfsetbuttcap%
\pgfsetroundjoin%
\definecolor{currentfill}{rgb}{0.000000,0.000000,0.000000}%
\pgfsetfillcolor{currentfill}%
\pgfsetlinewidth{0.803000pt}%
\definecolor{currentstroke}{rgb}{0.000000,0.000000,0.000000}%
\pgfsetstrokecolor{currentstroke}%
\pgfsetdash{}{0pt}%
\pgfsys@defobject{currentmarker}{\pgfqpoint{0.000000in}{-0.048611in}}{\pgfqpoint{0.000000in}{0.000000in}}{%
\pgfpathmoveto{\pgfqpoint{0.000000in}{0.000000in}}%
\pgfpathlineto{\pgfqpoint{0.000000in}{-0.048611in}}%
\pgfusepath{stroke,fill}%
}%
\begin{pgfscope}%
\pgfsys@transformshift{1.824387in}{0.502778in}%
\pgfsys@useobject{currentmarker}{}%
\end{pgfscope}%
\end{pgfscope}%
\begin{pgfscope}%
\definecolor{textcolor}{rgb}{0.000000,0.000000,0.000000}%
\pgfsetstrokecolor{textcolor}%
\pgfsetfillcolor{textcolor}%
\pgftext[x=1.824387in,y=0.405556in,,top]{\color{textcolor}\rmfamily\fontsize{8.000000}{9.600000}\selectfont 200}%
\end{pgfscope}%
\begin{pgfscope}%
\pgfpathrectangle{\pgfqpoint{0.559653in}{0.502778in}}{\pgfqpoint{2.529469in}{1.877222in}}%
\pgfusepath{clip}%
\pgfsetbuttcap%
\pgfsetroundjoin%
\pgfsetlinewidth{0.501875pt}%
\definecolor{currentstroke}{rgb}{0.000000,0.000000,0.000000}%
\pgfsetstrokecolor{currentstroke}%
\pgfsetstrokeopacity{0.500000}%
\pgfsetdash{{1.850000pt}{0.800000pt}}{0.000000pt}%
\pgfpathmoveto{\pgfqpoint{2.449256in}{0.502778in}}%
\pgfpathlineto{\pgfqpoint{2.449256in}{2.380000in}}%
\pgfusepath{stroke}%
\end{pgfscope}%
\begin{pgfscope}%
\pgfsetbuttcap%
\pgfsetroundjoin%
\definecolor{currentfill}{rgb}{0.000000,0.000000,0.000000}%
\pgfsetfillcolor{currentfill}%
\pgfsetlinewidth{0.803000pt}%
\definecolor{currentstroke}{rgb}{0.000000,0.000000,0.000000}%
\pgfsetstrokecolor{currentstroke}%
\pgfsetdash{}{0pt}%
\pgfsys@defobject{currentmarker}{\pgfqpoint{0.000000in}{-0.048611in}}{\pgfqpoint{0.000000in}{0.000000in}}{%
\pgfpathmoveto{\pgfqpoint{0.000000in}{0.000000in}}%
\pgfpathlineto{\pgfqpoint{0.000000in}{-0.048611in}}%
\pgfusepath{stroke,fill}%
}%
\begin{pgfscope}%
\pgfsys@transformshift{2.449256in}{0.502778in}%
\pgfsys@useobject{currentmarker}{}%
\end{pgfscope}%
\end{pgfscope}%
\begin{pgfscope}%
\definecolor{textcolor}{rgb}{0.000000,0.000000,0.000000}%
\pgfsetstrokecolor{textcolor}%
\pgfsetfillcolor{textcolor}%
\pgftext[x=2.449256in,y=0.405556in,,top]{\color{textcolor}\rmfamily\fontsize{8.000000}{9.600000}\selectfont 300}%
\end{pgfscope}%
\begin{pgfscope}%
\pgfpathrectangle{\pgfqpoint{0.559653in}{0.502778in}}{\pgfqpoint{2.529469in}{1.877222in}}%
\pgfusepath{clip}%
\pgfsetbuttcap%
\pgfsetroundjoin%
\pgfsetlinewidth{0.501875pt}%
\definecolor{currentstroke}{rgb}{0.000000,0.000000,0.000000}%
\pgfsetstrokecolor{currentstroke}%
\pgfsetstrokeopacity{0.500000}%
\pgfsetdash{{1.850000pt}{0.800000pt}}{0.000000pt}%
\pgfpathmoveto{\pgfqpoint{3.074125in}{0.502778in}}%
\pgfpathlineto{\pgfqpoint{3.074125in}{2.380000in}}%
\pgfusepath{stroke}%
\end{pgfscope}%
\begin{pgfscope}%
\pgfsetbuttcap%
\pgfsetroundjoin%
\definecolor{currentfill}{rgb}{0.000000,0.000000,0.000000}%
\pgfsetfillcolor{currentfill}%
\pgfsetlinewidth{0.803000pt}%
\definecolor{currentstroke}{rgb}{0.000000,0.000000,0.000000}%
\pgfsetstrokecolor{currentstroke}%
\pgfsetdash{}{0pt}%
\pgfsys@defobject{currentmarker}{\pgfqpoint{0.000000in}{-0.048611in}}{\pgfqpoint{0.000000in}{0.000000in}}{%
\pgfpathmoveto{\pgfqpoint{0.000000in}{0.000000in}}%
\pgfpathlineto{\pgfqpoint{0.000000in}{-0.048611in}}%
\pgfusepath{stroke,fill}%
}%
\begin{pgfscope}%
\pgfsys@transformshift{3.074125in}{0.502778in}%
\pgfsys@useobject{currentmarker}{}%
\end{pgfscope}%
\end{pgfscope}%
\begin{pgfscope}%
\definecolor{textcolor}{rgb}{0.000000,0.000000,0.000000}%
\pgfsetstrokecolor{textcolor}%
\pgfsetfillcolor{textcolor}%
\pgftext[x=3.074125in,y=0.405556in,,top]{\color{textcolor}\rmfamily\fontsize{8.000000}{9.600000}\selectfont 400}%
\end{pgfscope}%
\begin{pgfscope}%
\definecolor{textcolor}{rgb}{0.000000,0.000000,0.000000}%
\pgfsetstrokecolor{textcolor}%
\pgfsetfillcolor{textcolor}%
\pgftext[x=1.824387in,y=0.251234in,,top]{\color{textcolor}\rmfamily\fontsize{8.000000}{9.600000}\selectfont \(\displaystyle |\Omega|\)}%
\end{pgfscope}%
\begin{pgfscope}%
\pgfpathrectangle{\pgfqpoint{0.559653in}{0.502778in}}{\pgfqpoint{2.529469in}{1.877222in}}%
\pgfusepath{clip}%
\pgfsetbuttcap%
\pgfsetroundjoin%
\pgfsetlinewidth{0.501875pt}%
\definecolor{currentstroke}{rgb}{0.000000,0.000000,0.000000}%
\pgfsetstrokecolor{currentstroke}%
\pgfsetstrokeopacity{0.500000}%
\pgfsetdash{{1.850000pt}{0.800000pt}}{0.000000pt}%
\pgfpathmoveto{\pgfqpoint{0.559653in}{0.562936in}}%
\pgfpathlineto{\pgfqpoint{3.089121in}{0.562936in}}%
\pgfusepath{stroke}%
\end{pgfscope}%
\begin{pgfscope}%
\pgfsetbuttcap%
\pgfsetroundjoin%
\definecolor{currentfill}{rgb}{0.000000,0.000000,0.000000}%
\pgfsetfillcolor{currentfill}%
\pgfsetlinewidth{0.803000pt}%
\definecolor{currentstroke}{rgb}{0.000000,0.000000,0.000000}%
\pgfsetstrokecolor{currentstroke}%
\pgfsetdash{}{0pt}%
\pgfsys@defobject{currentmarker}{\pgfqpoint{-0.048611in}{0.000000in}}{\pgfqpoint{0.000000in}{0.000000in}}{%
\pgfpathmoveto{\pgfqpoint{0.000000in}{0.000000in}}%
\pgfpathlineto{\pgfqpoint{-0.048611in}{0.000000in}}%
\pgfusepath{stroke,fill}%
}%
\begin{pgfscope}%
\pgfsys@transformshift{0.559653in}{0.562936in}%
\pgfsys@useobject{currentmarker}{}%
\end{pgfscope}%
\end{pgfscope}%
\begin{pgfscope}%
\definecolor{textcolor}{rgb}{0.000000,0.000000,0.000000}%
\pgfsetstrokecolor{textcolor}%
\pgfsetfillcolor{textcolor}%
\pgftext[x=0.311580in, y=0.524356in, left, base]{\color{textcolor}\rmfamily\fontsize{8.000000}{9.600000}\selectfont 0.0}%
\end{pgfscope}%
\begin{pgfscope}%
\pgfpathrectangle{\pgfqpoint{0.559653in}{0.502778in}}{\pgfqpoint{2.529469in}{1.877222in}}%
\pgfusepath{clip}%
\pgfsetbuttcap%
\pgfsetroundjoin%
\pgfsetlinewidth{0.501875pt}%
\definecolor{currentstroke}{rgb}{0.000000,0.000000,0.000000}%
\pgfsetstrokecolor{currentstroke}%
\pgfsetstrokeopacity{0.500000}%
\pgfsetdash{{1.850000pt}{0.800000pt}}{0.000000pt}%
\pgfpathmoveto{\pgfqpoint{0.559653in}{0.887871in}}%
\pgfpathlineto{\pgfqpoint{3.089121in}{0.887871in}}%
\pgfusepath{stroke}%
\end{pgfscope}%
\begin{pgfscope}%
\pgfsetbuttcap%
\pgfsetroundjoin%
\definecolor{currentfill}{rgb}{0.000000,0.000000,0.000000}%
\pgfsetfillcolor{currentfill}%
\pgfsetlinewidth{0.803000pt}%
\definecolor{currentstroke}{rgb}{0.000000,0.000000,0.000000}%
\pgfsetstrokecolor{currentstroke}%
\pgfsetdash{}{0pt}%
\pgfsys@defobject{currentmarker}{\pgfqpoint{-0.048611in}{0.000000in}}{\pgfqpoint{0.000000in}{0.000000in}}{%
\pgfpathmoveto{\pgfqpoint{0.000000in}{0.000000in}}%
\pgfpathlineto{\pgfqpoint{-0.048611in}{0.000000in}}%
\pgfusepath{stroke,fill}%
}%
\begin{pgfscope}%
\pgfsys@transformshift{0.559653in}{0.887871in}%
\pgfsys@useobject{currentmarker}{}%
\end{pgfscope}%
\end{pgfscope}%
\begin{pgfscope}%
\definecolor{textcolor}{rgb}{0.000000,0.000000,0.000000}%
\pgfsetstrokecolor{textcolor}%
\pgfsetfillcolor{textcolor}%
\pgftext[x=0.311580in, y=0.849290in, left, base]{\color{textcolor}\rmfamily\fontsize{8.000000}{9.600000}\selectfont 0.1}%
\end{pgfscope}%
\begin{pgfscope}%
\pgfpathrectangle{\pgfqpoint{0.559653in}{0.502778in}}{\pgfqpoint{2.529469in}{1.877222in}}%
\pgfusepath{clip}%
\pgfsetbuttcap%
\pgfsetroundjoin%
\pgfsetlinewidth{0.501875pt}%
\definecolor{currentstroke}{rgb}{0.000000,0.000000,0.000000}%
\pgfsetstrokecolor{currentstroke}%
\pgfsetstrokeopacity{0.500000}%
\pgfsetdash{{1.850000pt}{0.800000pt}}{0.000000pt}%
\pgfpathmoveto{\pgfqpoint{0.559653in}{1.212805in}}%
\pgfpathlineto{\pgfqpoint{3.089121in}{1.212805in}}%
\pgfusepath{stroke}%
\end{pgfscope}%
\begin{pgfscope}%
\pgfsetbuttcap%
\pgfsetroundjoin%
\definecolor{currentfill}{rgb}{0.000000,0.000000,0.000000}%
\pgfsetfillcolor{currentfill}%
\pgfsetlinewidth{0.803000pt}%
\definecolor{currentstroke}{rgb}{0.000000,0.000000,0.000000}%
\pgfsetstrokecolor{currentstroke}%
\pgfsetdash{}{0pt}%
\pgfsys@defobject{currentmarker}{\pgfqpoint{-0.048611in}{0.000000in}}{\pgfqpoint{0.000000in}{0.000000in}}{%
\pgfpathmoveto{\pgfqpoint{0.000000in}{0.000000in}}%
\pgfpathlineto{\pgfqpoint{-0.048611in}{0.000000in}}%
\pgfusepath{stroke,fill}%
}%
\begin{pgfscope}%
\pgfsys@transformshift{0.559653in}{1.212805in}%
\pgfsys@useobject{currentmarker}{}%
\end{pgfscope}%
\end{pgfscope}%
\begin{pgfscope}%
\definecolor{textcolor}{rgb}{0.000000,0.000000,0.000000}%
\pgfsetstrokecolor{textcolor}%
\pgfsetfillcolor{textcolor}%
\pgftext[x=0.311580in, y=1.174225in, left, base]{\color{textcolor}\rmfamily\fontsize{8.000000}{9.600000}\selectfont 0.2}%
\end{pgfscope}%
\begin{pgfscope}%
\pgfpathrectangle{\pgfqpoint{0.559653in}{0.502778in}}{\pgfqpoint{2.529469in}{1.877222in}}%
\pgfusepath{clip}%
\pgfsetbuttcap%
\pgfsetroundjoin%
\pgfsetlinewidth{0.501875pt}%
\definecolor{currentstroke}{rgb}{0.000000,0.000000,0.000000}%
\pgfsetstrokecolor{currentstroke}%
\pgfsetstrokeopacity{0.500000}%
\pgfsetdash{{1.850000pt}{0.800000pt}}{0.000000pt}%
\pgfpathmoveto{\pgfqpoint{0.559653in}{1.537740in}}%
\pgfpathlineto{\pgfqpoint{3.089121in}{1.537740in}}%
\pgfusepath{stroke}%
\end{pgfscope}%
\begin{pgfscope}%
\pgfsetbuttcap%
\pgfsetroundjoin%
\definecolor{currentfill}{rgb}{0.000000,0.000000,0.000000}%
\pgfsetfillcolor{currentfill}%
\pgfsetlinewidth{0.803000pt}%
\definecolor{currentstroke}{rgb}{0.000000,0.000000,0.000000}%
\pgfsetstrokecolor{currentstroke}%
\pgfsetdash{}{0pt}%
\pgfsys@defobject{currentmarker}{\pgfqpoint{-0.048611in}{0.000000in}}{\pgfqpoint{0.000000in}{0.000000in}}{%
\pgfpathmoveto{\pgfqpoint{0.000000in}{0.000000in}}%
\pgfpathlineto{\pgfqpoint{-0.048611in}{0.000000in}}%
\pgfusepath{stroke,fill}%
}%
\begin{pgfscope}%
\pgfsys@transformshift{0.559653in}{1.537740in}%
\pgfsys@useobject{currentmarker}{}%
\end{pgfscope}%
\end{pgfscope}%
\begin{pgfscope}%
\definecolor{textcolor}{rgb}{0.000000,0.000000,0.000000}%
\pgfsetstrokecolor{textcolor}%
\pgfsetfillcolor{textcolor}%
\pgftext[x=0.311580in, y=1.499160in, left, base]{\color{textcolor}\rmfamily\fontsize{8.000000}{9.600000}\selectfont 0.3}%
\end{pgfscope}%
\begin{pgfscope}%
\pgfpathrectangle{\pgfqpoint{0.559653in}{0.502778in}}{\pgfqpoint{2.529469in}{1.877222in}}%
\pgfusepath{clip}%
\pgfsetbuttcap%
\pgfsetroundjoin%
\pgfsetlinewidth{0.501875pt}%
\definecolor{currentstroke}{rgb}{0.000000,0.000000,0.000000}%
\pgfsetstrokecolor{currentstroke}%
\pgfsetstrokeopacity{0.500000}%
\pgfsetdash{{1.850000pt}{0.800000pt}}{0.000000pt}%
\pgfpathmoveto{\pgfqpoint{0.559653in}{1.862674in}}%
\pgfpathlineto{\pgfqpoint{3.089121in}{1.862674in}}%
\pgfusepath{stroke}%
\end{pgfscope}%
\begin{pgfscope}%
\pgfsetbuttcap%
\pgfsetroundjoin%
\definecolor{currentfill}{rgb}{0.000000,0.000000,0.000000}%
\pgfsetfillcolor{currentfill}%
\pgfsetlinewidth{0.803000pt}%
\definecolor{currentstroke}{rgb}{0.000000,0.000000,0.000000}%
\pgfsetstrokecolor{currentstroke}%
\pgfsetdash{}{0pt}%
\pgfsys@defobject{currentmarker}{\pgfqpoint{-0.048611in}{0.000000in}}{\pgfqpoint{0.000000in}{0.000000in}}{%
\pgfpathmoveto{\pgfqpoint{0.000000in}{0.000000in}}%
\pgfpathlineto{\pgfqpoint{-0.048611in}{0.000000in}}%
\pgfusepath{stroke,fill}%
}%
\begin{pgfscope}%
\pgfsys@transformshift{0.559653in}{1.862674in}%
\pgfsys@useobject{currentmarker}{}%
\end{pgfscope}%
\end{pgfscope}%
\begin{pgfscope}%
\definecolor{textcolor}{rgb}{0.000000,0.000000,0.000000}%
\pgfsetstrokecolor{textcolor}%
\pgfsetfillcolor{textcolor}%
\pgftext[x=0.311580in, y=1.824094in, left, base]{\color{textcolor}\rmfamily\fontsize{8.000000}{9.600000}\selectfont 0.4}%
\end{pgfscope}%
\begin{pgfscope}%
\pgfpathrectangle{\pgfqpoint{0.559653in}{0.502778in}}{\pgfqpoint{2.529469in}{1.877222in}}%
\pgfusepath{clip}%
\pgfsetbuttcap%
\pgfsetroundjoin%
\pgfsetlinewidth{0.501875pt}%
\definecolor{currentstroke}{rgb}{0.000000,0.000000,0.000000}%
\pgfsetstrokecolor{currentstroke}%
\pgfsetstrokeopacity{0.500000}%
\pgfsetdash{{1.850000pt}{0.800000pt}}{0.000000pt}%
\pgfpathmoveto{\pgfqpoint{0.559653in}{2.187609in}}%
\pgfpathlineto{\pgfqpoint{3.089121in}{2.187609in}}%
\pgfusepath{stroke}%
\end{pgfscope}%
\begin{pgfscope}%
\pgfsetbuttcap%
\pgfsetroundjoin%
\definecolor{currentfill}{rgb}{0.000000,0.000000,0.000000}%
\pgfsetfillcolor{currentfill}%
\pgfsetlinewidth{0.803000pt}%
\definecolor{currentstroke}{rgb}{0.000000,0.000000,0.000000}%
\pgfsetstrokecolor{currentstroke}%
\pgfsetdash{}{0pt}%
\pgfsys@defobject{currentmarker}{\pgfqpoint{-0.048611in}{0.000000in}}{\pgfqpoint{0.000000in}{0.000000in}}{%
\pgfpathmoveto{\pgfqpoint{0.000000in}{0.000000in}}%
\pgfpathlineto{\pgfqpoint{-0.048611in}{0.000000in}}%
\pgfusepath{stroke,fill}%
}%
\begin{pgfscope}%
\pgfsys@transformshift{0.559653in}{2.187609in}%
\pgfsys@useobject{currentmarker}{}%
\end{pgfscope}%
\end{pgfscope}%
\begin{pgfscope}%
\definecolor{textcolor}{rgb}{0.000000,0.000000,0.000000}%
\pgfsetstrokecolor{textcolor}%
\pgfsetfillcolor{textcolor}%
\pgftext[x=0.311580in, y=2.149029in, left, base]{\color{textcolor}\rmfamily\fontsize{8.000000}{9.600000}\selectfont 0.5}%
\end{pgfscope}%
\begin{pgfscope}%
\definecolor{textcolor}{rgb}{0.000000,0.000000,0.000000}%
\pgfsetstrokecolor{textcolor}%
\pgfsetfillcolor{textcolor}%
\pgftext[x=0.256024in,y=1.441389in,,bottom,rotate=90.000000]{\color{textcolor}\rmfamily\fontsize{8.000000}{9.600000}\selectfont Average runtime [ms]}%
\end{pgfscope}%
\begin{pgfscope}%
\pgfpathrectangle{\pgfqpoint{0.559653in}{0.502778in}}{\pgfqpoint{2.529469in}{1.877222in}}%
\pgfusepath{clip}%
\pgfsetrectcap%
\pgfsetroundjoin%
\pgfsetlinewidth{1.505625pt}%
\definecolor{currentstroke}{rgb}{0.000000,0.000000,0.000000}%
\pgfsetstrokecolor{currentstroke}%
\pgfsetdash{}{0pt}%
\pgfpathmoveto{\pgfqpoint{0.674629in}{0.690855in}}%
\pgfpathlineto{\pgfqpoint{0.724618in}{0.742373in}}%
\pgfpathlineto{\pgfqpoint{0.774608in}{0.790676in}}%
\pgfpathlineto{\pgfqpoint{0.874587in}{0.910197in}}%
\pgfpathlineto{\pgfqpoint{0.974566in}{0.937839in}}%
\pgfpathlineto{\pgfqpoint{1.174524in}{1.114928in}}%
\pgfpathlineto{\pgfqpoint{1.374482in}{1.326709in}}%
\pgfpathlineto{\pgfqpoint{1.774398in}{1.676698in}}%
\pgfpathlineto{\pgfqpoint{2.174314in}{1.673823in}}%
\pgfpathlineto{\pgfqpoint{2.974146in}{2.294672in}}%
\pgfusepath{stroke}%
\end{pgfscope}%
\begin{pgfscope}%
\pgfpathrectangle{\pgfqpoint{0.559653in}{0.502778in}}{\pgfqpoint{2.529469in}{1.877222in}}%
\pgfusepath{clip}%
\pgfsetbuttcap%
\pgfsetroundjoin%
\pgfsetlinewidth{1.505625pt}%
\definecolor{currentstroke}{rgb}{1.000000,0.000000,0.000000}%
\pgfsetstrokecolor{currentstroke}%
\pgfsetdash{{5.550000pt}{2.400000pt}}{0.000000pt}%
\pgfpathmoveto{\pgfqpoint{0.674629in}{0.588106in}}%
\pgfpathlineto{\pgfqpoint{0.724618in}{0.596643in}}%
\pgfpathlineto{\pgfqpoint{0.774608in}{0.607404in}}%
\pgfpathlineto{\pgfqpoint{0.874587in}{0.629049in}}%
\pgfpathlineto{\pgfqpoint{0.974566in}{0.632078in}}%
\pgfpathlineto{\pgfqpoint{1.174524in}{0.666522in}}%
\pgfpathlineto{\pgfqpoint{1.374482in}{0.702491in}}%
\pgfpathlineto{\pgfqpoint{1.774398in}{0.749787in}}%
\pgfpathlineto{\pgfqpoint{2.174314in}{0.747734in}}%
\pgfpathlineto{\pgfqpoint{2.974146in}{0.913412in}}%
\pgfusepath{stroke}%
\end{pgfscope}%
\begin{pgfscope}%
\pgfsetrectcap%
\pgfsetmiterjoin%
\pgfsetlinewidth{0.803000pt}%
\definecolor{currentstroke}{rgb}{0.000000,0.000000,0.000000}%
\pgfsetstrokecolor{currentstroke}%
\pgfsetdash{}{0pt}%
\pgfpathmoveto{\pgfqpoint{0.559653in}{0.502778in}}%
\pgfpathlineto{\pgfqpoint{0.559653in}{2.380000in}}%
\pgfusepath{stroke}%
\end{pgfscope}%
\begin{pgfscope}%
\pgfsetrectcap%
\pgfsetmiterjoin%
\pgfsetlinewidth{0.803000pt}%
\definecolor{currentstroke}{rgb}{0.000000,0.000000,0.000000}%
\pgfsetstrokecolor{currentstroke}%
\pgfsetdash{}{0pt}%
\pgfpathmoveto{\pgfqpoint{3.089121in}{0.502778in}}%
\pgfpathlineto{\pgfqpoint{3.089121in}{2.380000in}}%
\pgfusepath{stroke}%
\end{pgfscope}%
\begin{pgfscope}%
\pgfsetrectcap%
\pgfsetmiterjoin%
\pgfsetlinewidth{0.803000pt}%
\definecolor{currentstroke}{rgb}{0.000000,0.000000,0.000000}%
\pgfsetstrokecolor{currentstroke}%
\pgfsetdash{}{0pt}%
\pgfpathmoveto{\pgfqpoint{0.559653in}{0.502778in}}%
\pgfpathlineto{\pgfqpoint{3.089121in}{0.502778in}}%
\pgfusepath{stroke}%
\end{pgfscope}%
\begin{pgfscope}%
\pgfsetrectcap%
\pgfsetmiterjoin%
\pgfsetlinewidth{0.803000pt}%
\definecolor{currentstroke}{rgb}{0.000000,0.000000,0.000000}%
\pgfsetstrokecolor{currentstroke}%
\pgfsetdash{}{0pt}%
\pgfpathmoveto{\pgfqpoint{0.559653in}{2.380000in}}%
\pgfpathlineto{\pgfqpoint{3.089121in}{2.380000in}}%
\pgfusepath{stroke}%
\end{pgfscope}%
\begin{pgfscope}%
\pgfsetbuttcap%
\pgfsetmiterjoin%
\definecolor{currentfill}{rgb}{1.000000,1.000000,1.000000}%
\pgfsetfillcolor{currentfill}%
\pgfsetlinewidth{1.003750pt}%
\definecolor{currentstroke}{rgb}{0.800000,0.800000,0.800000}%
\pgfsetstrokecolor{currentstroke}%
\pgfsetdash{}{0pt}%
\pgfpathmoveto{\pgfqpoint{0.637431in}{1.981234in}}%
\pgfpathlineto{\pgfqpoint{1.845815in}{1.981234in}}%
\pgfpathquadraticcurveto{\pgfqpoint{1.868037in}{1.981234in}}{\pgfqpoint{1.868037in}{2.003457in}}%
\pgfpathlineto{\pgfqpoint{1.868037in}{2.302222in}}%
\pgfpathquadraticcurveto{\pgfqpoint{1.868037in}{2.324444in}}{\pgfqpoint{1.845815in}{2.324444in}}%
\pgfpathlineto{\pgfqpoint{0.637431in}{2.324444in}}%
\pgfpathquadraticcurveto{\pgfqpoint{0.615208in}{2.324444in}}{\pgfqpoint{0.615208in}{2.302222in}}%
\pgfpathlineto{\pgfqpoint{0.615208in}{2.003457in}}%
\pgfpathquadraticcurveto{\pgfqpoint{0.615208in}{1.981234in}}{\pgfqpoint{0.637431in}{1.981234in}}%
\pgfpathclose%
\pgfusepath{stroke,fill}%
\end{pgfscope}%
\begin{pgfscope}%
\pgfsetrectcap%
\pgfsetroundjoin%
\pgfsetlinewidth{1.505625pt}%
\definecolor{currentstroke}{rgb}{0.000000,0.000000,0.000000}%
\pgfsetstrokecolor{currentstroke}%
\pgfsetdash{}{0pt}%
\pgfpathmoveto{\pgfqpoint{0.659653in}{2.241111in}}%
\pgfpathlineto{\pgfqpoint{0.881875in}{2.241111in}}%
\pgfusepath{stroke}%
\end{pgfscope}%
\begin{pgfscope}%
\definecolor{textcolor}{rgb}{0.000000,0.000000,0.000000}%
\pgfsetstrokecolor{textcolor}%
\pgfsetfillcolor{textcolor}%
\pgftext[x=0.970764in,y=2.202222in,left,base]{\color{textcolor}\rmfamily\fontsize{8.000000}{9.600000}\selectfont DisjointSet}%
\end{pgfscope}%
\begin{pgfscope}%
\pgfsetbuttcap%
\pgfsetroundjoin%
\pgfsetlinewidth{1.505625pt}%
\definecolor{currentstroke}{rgb}{1.000000,0.000000,0.000000}%
\pgfsetstrokecolor{currentstroke}%
\pgfsetdash{{5.550000pt}{2.400000pt}}{0.000000pt}%
\pgfpathmoveto{\pgfqpoint{0.659653in}{2.086173in}}%
\pgfpathlineto{\pgfqpoint{0.881875in}{2.086173in}}%
\pgfusepath{stroke}%
\end{pgfscope}%
\begin{pgfscope}%
\definecolor{textcolor}{rgb}{0.000000,0.000000,0.000000}%
\pgfsetstrokecolor{textcolor}%
\pgfsetfillcolor{textcolor}%
\pgftext[x=0.970764in,y=2.047284in,left,base]{\color{textcolor}\rmfamily\fontsize{8.000000}{9.600000}\selectfont Cached operator}%
\end{pgfscope}%
\end{pgfpicture}%
\makeatother%
\endgroup%